\documentclass{article}
\usepackage[utf8]{inputenc}
\usepackage{amsmath, amsfonts, amsthm}
\usepackage{graphicx}
\usepackage{multirow}
\usepackage[titletoc]{appendix}
\usepackage{nicematrix}
\usepackage[
backend=biber,
style=numeric,
sorting=none, 
maxbibnames=99
]{biblatex}
\DeclareFieldFormat*{title}{#1}
\renewbibmacro{in:}{%
  \ifentrytype{article}
    {}
    {\bibstring{in}%
     \printunit{\intitlepunct}}}
\DefineBibliographyExtras{english}{%
}
\usepackage{svg}
\usepackage{duckuments}

\providecommand{\keywords}[1]
{
  \small	
  \textbf{\textit{Keywords---}} #1
}

\usepackage[margin=2cm]{geometry}
\usepackage{setspace}
\usepackage{appendix}
\usepackage{cases}
\usepackage{empheq}
\usepackage{authblk} 
\usepackage{enumitem}

\title{Travelling waves in a coarse-grained model of volume-filling \\ cell invasion: Simulations and comparisons}
\author[1]{Rebecca M. Crossley}
\author[1]{Philip K. Maini}
\author[2]{Tommaso Lorenzi}
\author[1]{Ruth E. Baker}

\affil[1]{Mathematical Institute, University of Oxford, Oxford OX2 6GG, UK.}
\affil[2]{Department of Mathematical Sciences ``G. L. Lagrange'', Politecnico di Torino, 10129 Torino, Italy}
\affil[ ]{\textit {rebecca.crossley@maths.ox.ac.uk, maini@maths.ox.ac.uk, tommaso.lorenzi@polito.it, baker@maths.ox.ac.uk}}

\date{}

\begin{document}
\maketitle
\setlength{\marginparwidth}{10pt}

\begin{abstract}
Many reaction-diffusion models produce travelling wave solutions that can be interpreted as waves of invasion in biological scenarios such as wound healing or tumour growth. 
These partial differential equation models have since been adapted to describe the interactions between cells and extracellular matrix (ECM), using a variety of different underlying assumptions.  
In this work, we derive a system of reaction-diffusion equations, with cross-species density-dependent diffusion, by coarse-graining an agent-based, volume-filling model of cell invasion into ECM. 
We study the resulting travelling wave solutions both numerically and analytically across various parameter regimes. 
Subsequently, we perform a systematic comparison between the behaviours observed in this model and those predicted by simpler models in the literature that do not take into account volume-filling effects in the same way. 
Our study justifies the use of some of these simpler, more analytically tractable models in reproducing the qualitative properties of the solutions in some parameter regimes, but it also reveals some interesting properties arising from the introduction of cell and ECM volume-filling effects, where standard model simplifications might not be appropriate.
\end{abstract}


\begin{section}{Introduction}
Cell invasion is central to a variety of biological phenomena, playing a key role in morphogenesis, tumour growth and tissue engineering. 
Many different mathematical approaches have been used to model cell invasion processes, from agent-based models describing the processes underlying invasion at a single-cell level, to partial differential equation (PDE) models that provide a cell population-level description of invasion in terms of cell density dynamics~\cite{giniunaite2020modelling}.
Whilst some of these PDE models have been formulated through adaptations of classical models for invasion processes in other biological contexts, many are derived by coarse-graining a cell-level description to produce PDEs that offer the corresponding population-level description. 
However, there remains a number of unanswered questions regarding the specific choice of model for a given application, including how varying assumptions in the agent-based model give rise to different PDE models, and to what extent these differences impact the cell population-level description~\cite{johnston2017co}. 

In this work, the impact of various modelling assumptions at the single-cell level is compared by investigating the qualitative and quantitative properties of the solutions of the resulting PDE models. 
In particular, since in many real-life instances of cell invasion the cells have to invade through extracellular matrix (ECM)~\cite{dallon_mathematical_1999, perumpanani1999extracellular, painter2003modelling} -- i.e. the network of proteins and other molecules that impact collective cell invasion by reducing space available for cells to migrate into -- the focus is on models of cell invasion into ECM.

The classic example of a model describing invasion of a single population is the Fisher-Kolmogorov-Pietrovskii-Piskunov (FKPP) model, 
which was first proposed in the context of the spread of an advantageous gene~\cite{fisher_wave_1937, kolmogorov1937study}. 
This model has seen a broad spectrum of applications in the natural sciences: most notably in cell biology~\cite{maini2004traveling, gerlee2016travelling} and ecology~\cite{okubo1989spatial, kot2001elements}, where travelling wave solutions are representative of invasion phenomena~\cite{canosa1973nonlinear, murray2002mathematical}. 
A model of cell invasion through ECM is presented in~\cite{el2021travelling}. 
This consists of a system of two coupled PDEs, with a non-linear cross-species density-dependent diffusion term and logistic growth, whereby proliferation of the cell population is limited by the presence of cells and ECM. 
A similar model is considered in~\cite{colson_travelling-wave_2021}, where proliferation depends only on the presence of cells. 
An obvious question to ask is how the predictions of such models may be affected by a consistent description of the role of volume-filling effects (i.e. cells and ECM take up some given volume, preventing cell invasion) across both proliferative and diffusive mechanisms of cell dynamics. 
In particular, the impact of crowding on cell motility is generally modelled at a population-level by the introduction of a density-dependent diffusion term, such as in~\cite{gurtin1977diffusion, sengers2007experimental, martin2010tumour, mcgillen2014general}. However these models provide a phenomenological description of the impact of crowding, rather than considering how interactions at the individual cell-level directly impact motility at the population-level.

This study aims to extend and apply the work in~\cite{simpson2009diffusing, simpson2009multi} to develop an agent-based model for cell invasion into ECM, taking into account volume-filling effects, where both cell motility and proliferation are impacted by the presence of other cells or ECM components. 
We make the simplifying assumption that space is the only factor limiting cell invasion, whereas other models in the literature \cite{ anderson2009microenvironment, klm2015hybrid} assume various other factors, such as nutrient-limited growth \cite{tam2018nutrient}. 
By coarse-graining this model, a limiting PDE description is formally derived and explored both analytically and numerically, making it possible to carry out a systematic comparison between the population-level behaviours observed in this model and those predicted by simpler models. 
In particular, we compare how the population-level behaviours predicted by this model relate to existing models built on different constitutive assumptions, such as the FKPP model, or the simpler models presented in~\cite{el2021travelling, colson_travelling-wave_2021}. Each of these simpler models can be recovered from the model presented in this work by neglecting specific terms, such as those capturing volume-filling effects.

\end{section}

\begin{section}{Mathematical model and preliminary results}
We begin by developing a simple one-dimensional, on-lattice, agent-based model of cell invasion into ECM that incorporates both cell motility and proliferation, and degradation of ECM, in the presence of volume-filling effects. We then coarse-grain this model to formally derive a corresponding PDE model that comprises a system of coupled PDEs for the densities of cells and ECM~\cite{simpson2009diffusing, bruna2012diffusion}.  

\subsection{Agent-based model}
In the simplified setting of this model, cells are represented as discrete agents that can proliferate and move on a one-dimensional uniform lattice, which constitutes the spatial domain, and can also degrade the surrounding ECM, which is regarded as being composed of discrete constitutive elements. The novel aspect of this model is the introduction of volume-filling effects, similar to the model described in~\cite{morris2020identifying}, which uses the methods described in~\cite{painter2002volume}, but extended to multiple populations~\cite{simpson2009multi}. 

Let the number of cells and ECM elements on lattice site $i=1,2,\ldots$ of width $\Delta$ at time $\tilde{t} \in \mathbb{R}^+$ of realisation $j = 1,2,\ldots, J$ of the model be denoted, respectively, by $u^j_i(\tilde{t})$ and $m^j_i(\tilde{t})$. 
We assume that ECM elements have constant density and are chosen to occupy a volume equal to that of a cell, such that at most $N$ cells or ECM elements can occupy each lattice site. 

The dynamics of the cells are governed by two mechanisms: proliferation, in which a cell places a daughter cell into the same lattice site it occupies; and motility, whereby cells can move to one of their two adjacent lattice sites. Moreover, ECM elements can be degraded by cells in the same lattice site as them. To incorporate volume-filling effects into the model, we prescribe that each lattice site has a maximum occupancy level $N$~\cite{taylor2016coupling} and assume that: 
\begin{enumerate}[label=(A\arabic*)]
\item if a cell attempts a move to a neighbouring lattice site, then the probability that the move is successful decreases linearly with the occupancy level of the target site, such that the probability of a successful move to a target site with occupancy level $N$ is zero; \label{Ass1}
\item if a cell attempts to proliferate, then the probability of success decreases linearly with the occupancy level of the site where the cell is located, such that the probability of a successful proliferation event in a site with occupancy level $N$ is zero. \label{Ass2}
\end{enumerate}

\paragraph{Probability of cell movement.} A cell attempts a movement in a time step $\tau$ with probability $p_{\rm m} \in[0,1]$, and the attempted movement from lattice site $i$ to either of the neighbouring lattice sites $i\pm1$ occurs with equal probability $1/2$. 
Using assumption~\ref{Ass1}, we can define the probability of movement to the left, $T_{i-}^{{{{\rm {m}}^j}}}(\tilde{t})$, or right, $T_{i+}^{{{\rm {m}}^j}}(\tilde{t})$, during the time interval $[\tilde{t}, \tilde{t}+\tau)$ of realisation $j$, as
\begin{equation}\label{def:Tm}
T_{i\pm}^{{{\rm {m}}^j}}(\tilde{t}) = \frac{p_{{\rm m}}}{2}\bigg(1-\frac{ u^j_{i\pm1}(\tilde{t})+ m^j_{i\pm1}(\tilde{t}) }{N}\bigg).
\end{equation}

\paragraph{Probability of cell proliferation.} A cell in lattice site $i$ attempts to proliferate in time step $\tau$ with probability $p_{\rm p} \in[0,1]$. 
If proliferation occurs, then the cell places a daughter cell into the same lattice site as itself. 
Using assumption~\ref{Ass2}, we can define the probability of proliferation, $T_i^{{{\rm {p}}^j}}(\tilde{t})$, during the time interval $[\tilde{t}, \tilde{t}+\tau)$ of realisation $j$, as
\begin{equation}\label{def:Tp}
T_i^{{{\rm {p}}^j}}(\tilde{t}) = p_{\rm p} \, \bigg(1- \frac{u^j_{i}(\tilde{t})+ m^j_{i}(\tilde{t}) }{N}\bigg).
\end{equation}

\medskip
Note that, the initial distributions of cells and ECM elements must be such that at most $N$ cells or ECM elements can occupy each lattice site to ensure the probabilities $T_{i\pm}^{{{\rm {m}}^j}}(\tilde{t}),\, T_i^{{{\rm {p}}^j}}(\tilde{t}) \geq0$ are well-defined. 
Under the assumption that the initial distributions of cells and ECM elements satisfy $0 \leq u^j_i(0) +  m^j_i(0) \leq N \; \text{for all} \;  j = 1,2,\ldots, J \; \text{and} \; i = 1,2,\ldots$, the definitions for the probabilities of cell movement and proliferation given by Equations~\eqref{def:Tm} and~\eqref{def:Tp} ensure that
\begin{equation}\label{eq:aprestAB}
0 \leq u^j_i(\tilde{t}) +  m^j_i(\tilde{t}) \leq N \;\; \text{for all} \;\;  j = 1,2,\ldots, J \;\; \text{and} \;\; i = 1,2,\ldots \;\; \text{for any} \;\; \tilde{t} \in \mathbb{R}^+.
\end{equation}

\paragraph{Probability of ECM degradation.} During the time interval  $[\tilde{t}, \tilde{t}+\tau)$ of realisation $j$, an element of ECM in lattice site $i$ is degraded by a cell on the same lattice site with probability $p_{\rm d}\in[0,1]$, such that the degradation per unit element of ECM, $T_i^{{{\rm {d}}^j}}(\tilde{t})$, is
\begin{equation}
T_i^{{{\rm {d}}^j}}(\tilde{t})=p_{\rm d} u_i^j(\tilde{t}). \nonumber
\end{equation}

\subsection{Corresponding coarse-grained model}
In order to derive a coarse-grained description of the agent-based model, we introduce the average occupancy of lattice site $i$ at time $\tilde{t}$ by cells and ECM elements over $J$ realisations of the model, denoted, respectively, by
$$
\langle u_i(\tilde{t}) \rangle = \frac{1}{J} \sum_{j=1}^J u^j_i(\tilde{t}) \;\;\;\; \text{and} \;\;\;\; \langle m_i(\tilde{t}) \rangle= \frac{1}{J} \sum_{j=1}^J m^j_i(\tilde{t}).
$$

\paragraph{Coarse-grained model of cell dynamics.}
We proceed to derive  a coarse-grained model by considering how the average occupancy in lattice site $i$ changes during the time interval $[\tilde{t}, \tilde{t}+\tau)$:
 \begin{align} \label{cell ABM eq}
    \langle u_i(\tilde{t}+\tau) \rangle &= \langle u_{i}(\tilde{t}) \rangle \nonumber\\ &\qquad+ \frac{p_{\rm m}}{2}\langle u_{i+1}(\tilde{t}) \rangle \bigg(1- \frac{ \langle u_{i}(\tilde{t})\rangle + \langle m_{i}(\tilde{t})\rangle }{N}\bigg)\nonumber\\&\qquad+\frac{p_{\rm m}}{2}\langle u_{i-1}(\tilde{t}) \rangle \bigg(1- \frac{ \langle u_{i}(\tilde{t})\rangle + \langle m_{i}(\tilde{t})\rangle }{N}\bigg) \nonumber\\&\qquad-\frac{p_{\rm m}}{2}\langle u_{i}(\tilde{t}) \rangle \bigg(1- \frac{\langle u_{i+1}(\tilde{t})\rangle+\langle m_{i+1}(\tilde{t})\rangle }{N}\bigg) \nonumber\\&\qquad-\frac{p_{\rm m}}{2}\langle u_{i}(\tilde{t}) \rangle \bigg(1- \frac{\langle u_{i-1}(\tilde{t})\rangle + \langle m_{i-1}(\tilde{t})\rangle }{N}\bigg) \nonumber\\ &\qquad+ p_{\rm p}\langle u_{i}(\tilde{t}) \rangle \bigg(1- \frac{\langle u_{i}(\tilde{t})\rangle + \langle m_{i}(\tilde{t})\rangle }{N}\bigg).
\end{align}
Note that, in writing down Equation~\eqref{cell ABM eq} we have used probabilistic approximations of the mean-field type which are frequently used for the coarse-graining of agent-based models and involve assuming independence of lattice sites (see, for example,~\cite{penington2011building}).
Rearranging Equation~\eqref{cell ABM eq} and dividing both sides by $\tau$ yields:
\begin{align} \label{cell ABM 2}
    \frac{\langle u_i(\tilde{t}+\tau) \rangle - \langle u_i(\tilde{t}) \rangle }{\tau}&= \frac{p_{\rm m} \Delta^2}{ 2\tau}\bigg[\frac{\langle u_{i-1}(\tilde{t}) \rangle -2 \langle u_{i}(\tilde{t}) \rangle +\langle u_{i+1}(\tilde{t}) \rangle}{ \Delta^2}\bigg]\nonumber \\ &\qquad+ \frac{p_{\rm m}\Delta^2}{2\tau N}\bigg[\frac{\langle u_{i}(\tilde{t}) \rangle (\langle m_{i-1}(\tilde{t}) \rangle -2\langle m_{i}(\tilde{t}) \rangle+\langle m_{i+1}(\tilde{t}) \rangle )}{\Delta^2}\bigg]\nonumber \\&\qquad-\frac{p_{\rm m}\Delta^2}{2\tau N}\bigg[\frac{\langle m_{i}(\tilde{t}) \rangle (\langle u_{i-1}(\tilde{t}) \rangle -2\langle u_{i}(\tilde{t}) \rangle +\langle u_{i+1}(\tilde{t}) \rangle)}{\Delta^2}\bigg] \nonumber \\&\qquad+\frac{p_{\rm p}}{\tau}\langle u_{i}(\tilde{t}) \rangle \bigg(1- \frac{\langle u_{i}(\tilde{t})\rangle + \langle m_{i}(\tilde{t})\rangle}{N}\bigg).
\end{align}
We now divide both sides of Equation~\eqref{cell ABM 2} by length scale $\Delta$, perform a Taylor expansion and take limits as $\Delta,\tau\to0$ to obtain a description of the cell density dynamics in terms of the variables $\tilde{u}(\tilde{x},\tilde{t})$ and $\tilde{m}(\tilde{x},\tilde{t})$, that are the continuum counterparts of $\langle u_i(\tilde{t}) \rangle/\Delta$ and $\langle m_i(\tilde{t}) \rangle/(\mu\Delta)$ that represent, respectively, the number density of cells and the density of ECM at position $\tilde{x} \in \mathbb{R}$ and time $\tilde{t}\in(0,\infty)$. The factor $\mu$ represents the number of cells equivalent to a unit mass of ECM and is introduced as a conversion factor between the density of ECM, as defined by mass of ECM per unit volume, and the number density of ECM elements, given by $\mu \tilde{m}(\tilde{x},\tilde{t})$. Under the assumptions
\begin{equation}
\lim_{\Delta, \tau\to0}\frac{p_{\rm m} \Delta ^2}{2\tau} = \tilde{D}, \qquad \lim_{\tau\to0}\frac{p_{\rm p}}{\tau}=\tilde{r}, \qquad \lim_{\Delta\to0}\frac{N}{\Delta}=\tilde{K}, \label{params}
\end{equation}
we obtain the following PDE for the cell density $\tilde{u}(\tilde{x},\tilde{t})$:
\begin{equation}
    \frac{\partial \tilde{u}}{\partial \tilde{t}}=\tilde{D}\frac{\partial}{\partial \tilde{x}}\bigg[\bigg(1-\frac{\tilde{u}+\tilde{\mu}\tilde{m}}{\tilde{K}}\bigg)\frac{\partial \tilde{u}}{\partial \tilde{x}}+\tilde{u}\frac{\partial}{\partial \tilde{x}}\bigg(\frac{\tilde{u}+\tilde{\mu}\tilde{m}}{\tilde{K}}\bigg)\bigg]+\tilde{r}\tilde{u}\bigg(1-\frac{\tilde{u}+\tilde{\mu}\tilde{m}}{\tilde{K}}\bigg), \label{dimu} \end{equation}
where $\tilde{x}\in\mathbb{R}$ and $\tilde{t}\in(0,\infty)$. 
Note that the first term on the right-hand side of Equation~\eqref{dimu} describes the movement of cells down gradients in cell density, with movement prevented by the presence of surrounding cells and ECM, as expected by the introduction of volume-filling effects. The second term models the motion of the cells down the ``total density gradient" of cells and ECM, $\tilde{u}+\tilde{\mu}\tilde{m}$. The third term captures cell proliferation, which is also impacted by volume-filling effects. From Equation~\eqref{dimu} it is clear that the parameter $\tilde{D}\geq0$, which is defined via Equation~\eqref{params}, can be regarded as the diffusion coefficient of the cells in the absence of ECM, while the parameters $\tilde{r}\geq0$ and $\tilde{K}>0$, which are also defined via Equation~\eqref{params}, are the intrinsic growth rate of the cell population, and the density corresponding to the maximum occupancy level (i.e. the carrying capacity), respectively. 

\paragraph{Coarse-grained model of ECM dynamics.} Probabilistic approximations similar to those underlying Equation~\eqref{cell ABM eq} give the following conservation equation for the evolution of ECM elements in lattice site $i$  during the time interval $[\tilde{t}, \tilde{t}+\tau)$: 
\begin{equation} \label{mABM}
    \langle m_{i}(\tilde{t}+\tau)\rangle = \langle m_{i}(\tilde{t})\rangle -p_{\rm d} \langle u_{i}(\tilde{t})\rangle\langle m_{i}(\tilde{t})\rangle .
\end{equation}
Rearranging Equation~\eqref{mABM}, dividing by $\Delta$ and $\tau$ and taking limits as $\Delta,\tau\to0$, under the assumption 
\begin{equation}
\lim_{\Delta, \tau\to0} \frac{p_{\rm d}\Delta}{\tau} = \tilde{\lambda}, \label{params2}
\end{equation}
we formally obtain the following differential equation for ECM density $\tilde{m}(\tilde{x},\tilde{t})$:
\begin{align}
    \frac{\partial \tilde{m}}{\partial \tilde{t}}=-\tilde{\lambda} \tilde{m} \tilde{u}, \label{dimm}
\end{align}
where  $\tilde{x} \in \mathbb{R}$ and $\tilde{t}\in(0,\infty)$. Here, the parameter $\tilde{\lambda} \geq 0$ defined via Equation~\eqref{params2} is the per cell degradation rate of ECM.

We observe that when there is no ECM degradation (i.e. if $\tilde{\lambda}=0$) and ECM is uniformly distributed at $\tilde{t}=0$ (i.e. if $\tilde{m}(\tilde{x},0) \equiv \tilde{m}^0$ where $\tilde{m}^0 \in \mathbb{R}^+$ with $0 \leq \tilde{m}^0 \leq \tilde{K}$), the mathematical model defined via Equations~\eqref{dimu} and~\eqref{dimm} simplifies to the following FKPP model of cell dynamics~\cite{fisher_wave_1937}:
\begin{equation}
    \label{FKPP}
    \frac{\partial \tilde{u}}{\partial \tilde{t}}=\hat{D}\frac{\partial ^2 \tilde{u}}{\partial \tilde{x}^2}+\hat{r}\tilde{u}\bigg(1-\frac{\tilde{u}}{\hat{K}}\bigg),
\end{equation}
where 
$$
\hat{D} = \left(1 - \dfrac{\tilde{\mu} \tilde{m}^0}{\tilde{K}} \right) \tilde{D}, \quad \hat{r} = \left(1 - \dfrac{\tilde{\mu} \tilde{m}^0}{\tilde{K}} \right) \tilde{r}, \quad \hat{K} = \left(1 - \dfrac{\tilde{\mu} \tilde{m}^0}{\tilde{K}} \right) \tilde{K}.
$$

\begin{subsection}{Non-dimensional coarse-grained model}
The mathematical model defined via Equations~\eqref{dimu} and~\eqref{dimm} can be non-dimensionalised by the introduction of the following non-dimensional variables:
\begin{equation*}
    u=\frac{\tilde{u}}{\tilde{K}},\quad    m=\frac{\tilde{\mu}\tilde{m}}{\tilde{K}}, \quad
    t=\tilde{t}\tilde{r}, \quad
    x=\sqrt{\frac{\tilde{r}}{\tilde{D}}}\tilde{x}, 
\end{equation*}
and written as:
\begin{align}{}
    \frac{\partial u}{\partial t}&=  \frac{\partial}{\partial x}\bigg[ (1- m)\frac{\partial u}{\partial x}+u\frac{\partial m}{\partial x}\bigg]+ u(1-u-m),  \label{NDeqn2_u} \\
    \frac{\partial m}{\partial t}&=-\lambda m u, \label{NDeqn2_m}
\end{align}
where $x\in\mathbb{R}$ and $t\in(0,\infty)$. 
Here, the only remaining parameter is $\lambda={\tilde{\lambda}\tilde{ K}}/{\tilde{r}}\geq0$ which is interpreted as the rescaled ECM degradation rate. 
We complement the model defined via Equations~\eqref{NDeqn2_u}-\eqref{NDeqn2_m} with no flux boundary conditions for Equation~\eqref{NDeqn2_u}:
\begin{equation}\label{NFBC}
(1- m)\frac{\partial u}{\partial x}+u\frac{\partial m}{\partial x} = 0 \bigg|_{x=0}, 
\end{equation}
and $u,\,{\partial u}/{\partial x}\to 0$ as $x\to\infty$. We also have the following initial conditions:
\begin{equation}\label{ass:ICPDE}
u(x,0)=u_0(x) \geq 0, \quad m(x,0)=m_0(x) \geq 0, \quad  0 \leq u_0(x) + m_0(x) \leq 1 \quad \forall \, x \in \mathbb{R}.
\end{equation}

We note that by assuming at the single-cell level that both the presence of cells and ECM elements impair the movement  and proliferation of the cells, the resulting population-level description for cell density evolution in Equation~\eqref{NDeqn2_u} exhibits a number of differences to similar models without volume-filling effects. 
For example, the model studied by El Hachem et al.~in \cite{el2021travelling} does not consider volume-filling of cells to impair cell movement, and therefore contains one less flux term, namely that accounting for movement of cells down the ``total density gradient''. 
This model can be recovered from Equation~\eqref{NDeqn2_u} by employing different underlying assumptions such that the probability of movement depends on the average available space (where space is only filled by ECM) between the target lattice site and the lattice site the cell occupies at time $\tilde{t}$.

\end{subsection}

\begin{subsection}{Numerical exploration of possible travelling wave solutions}
We are interested in the possible constant profile, constant speed travelling wave solutions displayed by the model defined via Equations~\eqref{NDeqn2_u}-\eqref{NDeqn2_m}.
As such, we first explore the range of possible behaviours numerically. 
We report on the results of numerical simulations carried out for the model posed on the spatial domain $(0,L)$, with $L>0$ sufficiently large so that the no flux boundary condition~\eqref{NFBC} at $x=L$ does not interact with the travelling wave. The simulations were subject to the following initial conditions: 
\begin{equation}
u(x,0)=\begin{cases} 1, \qquad & \text{if} \qquad x<\alpha, \\ 0 \qquad & \text{if} \qquad x\geq\alpha,  \label{hIC_u} \end{cases}
\end{equation}
\begin{equation}
m(x,0)=\begin{cases} 0, &\text{if} \qquad x<\alpha, \\ m_0 & \text{if} \qquad x\geq\alpha, \label{hIC_m} \end{cases} 
\end{equation}
where $0<\alpha\ll L$ represents the width of the initially invaded region at $t=0$ and $m_0\in[0,1)$ corresponds to the uninvaded density of ECM ahead of the cells. 

We note that, by design, the model~\eqref{NDeqn2_u}-\eqref{NDeqn2_m} does not permit travelling waves when there are initial conditions with compactly supported cell density and $m_0=1$. 
This is because cells require space ahead of the wave in order to invade; in any regions initially devoid of cells, the ECM cannot be degraded to allow cells to invade.
As such, we proceed by considering $m_0\in[0,1).$
Further specifics of the parameter values and the numerical methods used in this paper can be found in Appendix~\ref{appNS}.

\begin{figure}[h!]
    \centering
\hspace*{-.4cm} 
	\includegraphics[scale=0.575]{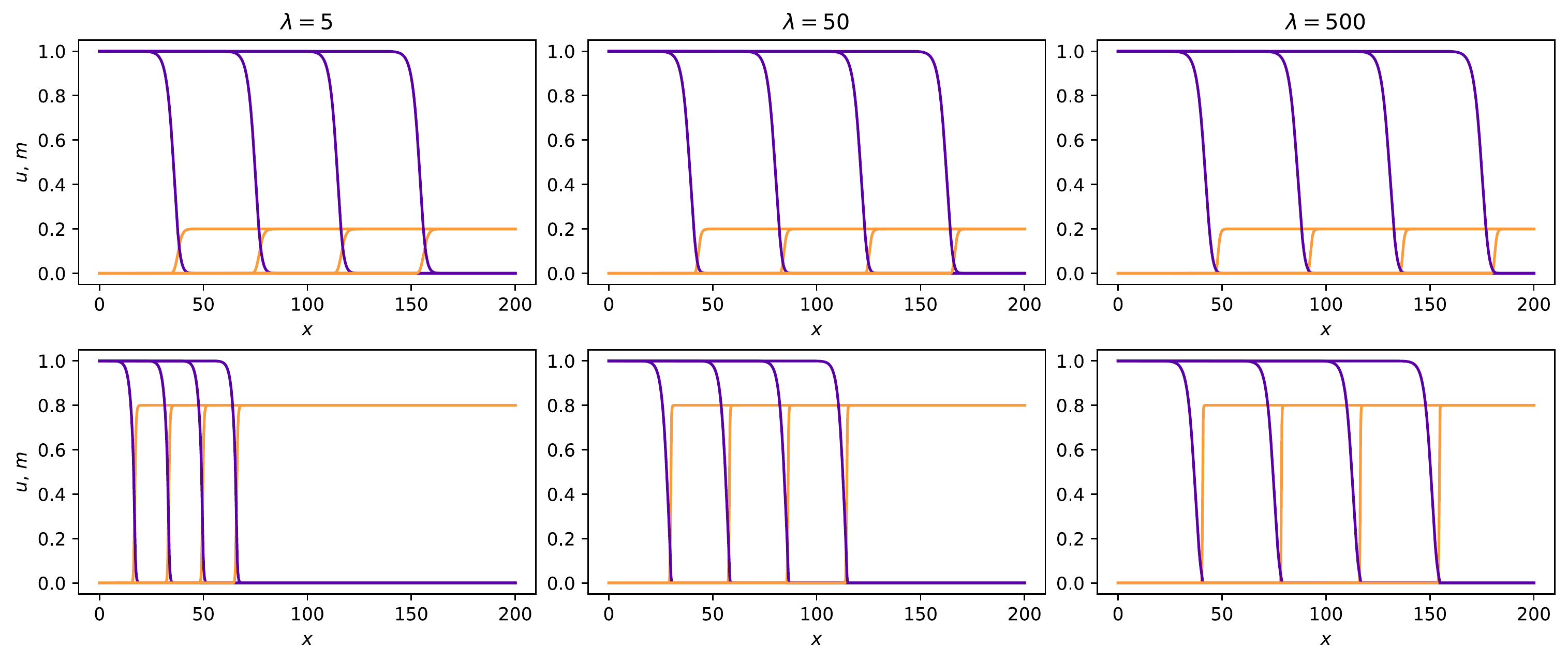}
    \caption{Numerical solutions of Equations~\eqref{NDeqn2_u}-\eqref{NDeqn2_m} subject to the initial conditions~\eqref{hIC_u}-\eqref{hIC_m}, for $m_0=0.2$ in the top row and $m_0=0.8$ in the bottom row, and for rescaled ECM degradation rates $\lambda=5,\,50,\,500.$ Cell densities are shown in purple and ECM densities in orange at times $t=25,\,50,\,75,\,100$ from left to right. Further specifics of the parameter values and the numerical methods used can be found in Appendix~\ref{appNS}.}
    \label{fig:miscTWprofile}
\end{figure}

As shown in Figure~\ref{fig:miscTWprofile}, when $m_0 \in [0,1)$ the solutions to Equations~\eqref{NDeqn2_u}-\eqref{NDeqn2_m} subject to the initial conditions~\eqref{hIC_u}-\eqref{hIC_m} converge to travelling waves whereby the cell density, $u$, decreases monotonically from one to zero and the ECM density, $m$, increases monotonically from zero to $m_0$.  
The numerical results in Figure~\ref{fig:miscTWprofile} also indicate that the speed of the travelling waves changes as the values of the parameters $\lambda$ and $m_0$ are changed. 
This is illustrated in more detail in Figure~\ref{fig:cvsdmvsm0}, that also shows that (in agreement with the analytical results presented in Section~\ref{TWA}) when $m_0 \in (0,1)$: if $\lambda \to 0^+$ then the speed of the travelling waves converges to $c=2 \, (1-m_0)$; whereas if $\lambda \to \infty$ then the speed of the travelling waves converges to $c=2.$ 

We also note that when $m_0=0$, the solutions to Equation~\eqref{NDeqn2_m} subject to the initial condition~\eqref{hIC_m} are such that $m(x,t) \equiv 0$ for all $t \geq 0$ and thus the model simplifies to the FKPP model~\eqref{FKPP} with $\hat{D}=\hat{r}=\hat{K}=1$, that is
\begin{equation}
    \label{FKPP1}
    \frac{\partial u}{\partial t}= \frac{\partial ^2 u}{\partial x^2}+ u \left(1-u\right).
\end{equation}
Consistent with this, numerical simulations indicate that when $m_0=0$, the cell density $u$ converges to a travelling wave that decreases monotonically from one to zero (results not shown), and travels with speed $c=2$ (i.e. the minimal  speed of travelling wave solutions to the FKPP model~\eqref{FKPP1}), see Figure~\ref{fig:cvsdmvsm0}. 

The numerical results summarised by Figure~\ref{fig:cvsdmvsm0} for $m_0 \in [0,1)$ show similar behaviours to that in~\cite{el2021travelling}, where no volume-filling effects of cells prevent cell movement, whilst a marked difference is observed for the case $m_0=1$, as discussed in Appendix~\ref{m0_1}. 

\begin{figure}[h!]
    \centering
	\includegraphics[scale=0.6]{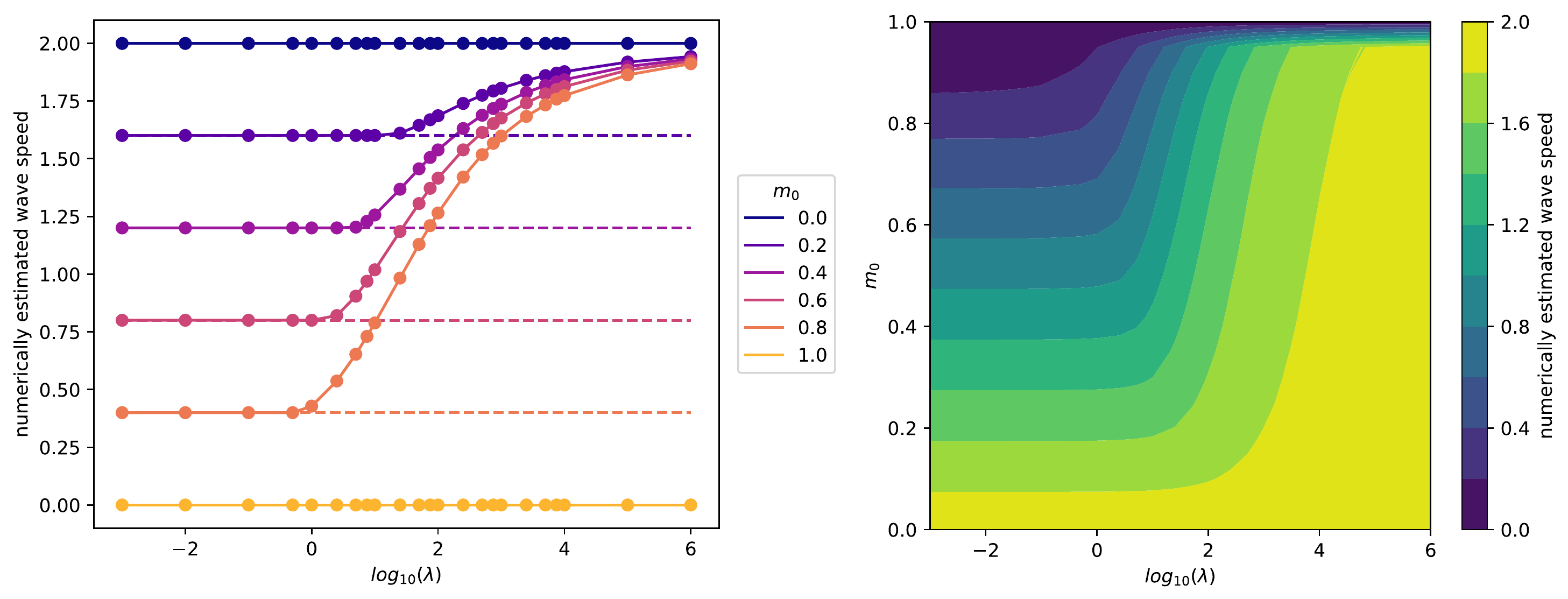}
    \caption{{The relationship between the numerically estimated speed (solid lines) of travelling wave solutions of Equations~\eqref{NDeqn2_u}-\eqref{NDeqn2_m} subject to the initial conditions~\eqref{hIC_u}-\eqref{hIC_m}. The dashed lines in the plot on the left highlight the value of $2(1-m_0)$. The numerically estimated travelling wave speed is obtained by tracing the point $X(t)$ such that $u(X(t), t)=0.1$. Further specifics of the parameter values and the numerical methods used can be found in Appendix~\ref{appNS}.}}
    \label{fig:cvsdmvsm0}
\end{figure}

\end{subsection}

\end{section}

\begin{section}{Travelling wave analysis}\label{TWA}
We seek travelling wave solutions of Equations~\eqref{NDeqn2_u}-\eqref{NDeqn2_m} by adopting the usual travelling wave ansatz $u(x,t)=U(z)$ and $m(x,t)=M(z)$ where $z=x-ct$ with $c>0$. 
Since numerical simulations indicate that, for our chosen initial conditions, travelling waves do not emerge when $m_0=1$ (see Appendix~\ref{m0_1}), we proceed with this study by exclusively considering the case where  $m_0\in[0,1)$ that gives
\begin{align}{}
    \frac{\mathrm{d}}{\mathrm{d} z}\bigg[(1-M) \frac{\mathrm{d} U}{\mathrm{d} z}+U \frac{\mathrm{d} M}{\mathrm{d} z}\bigg]+c\frac{\mathrm{d} U}{\mathrm{d} z}+U\big(1-U-M\big)&=0, \label{stdTW_u} \\
    c\frac{\mathrm{d} M}{\mathrm{d} z}-\lambda M U  &=0, \label{stdTW_m}
\end{align}
for $-\infty<z<\infty$ with boundary conditions
\begin{align}
    U(z)\to1 \quad &\text{as} \quad  z\to-\infty, \label{BCUz}\\
    U(z)\to0\quad  &\text{as}\quad  z\to\infty, \label{BCU2z} \\
    M(z)\to m_0 \quad &\text{as} \quad z\to\infty. \label{BCMz}
\end{align}
{By expanding Equation~\eqref{stdTW_u} and using Equation \eqref{stdTW_m} to substitute in $\mathrm{d}^2M/\mathrm{d}z^2$, we find that
\begin{equation}
\left(1 - M \right) \dfrac{{\rm d}^2 U}{{\rm d}z^2} + c \dfrac{{\rm d} U}{{\rm d}z} + U \big[\left(1 - M\right) - U \big] = -M \, \left[U \left(\dfrac{\lambda \, U}{c} \right)^2 + \dfrac{{\rm d} U}{{\rm d}z} \left(\dfrac{\lambda \, U}{c} \right)\right]. \label{subst_TW}
\end{equation}
}Equation~\eqref{stdTW_m}, subject to the boundary condition~\eqref{BCMz}, has a semi-explicit solution. 
That is, if $U(z)$ is known, then we can evaluate $M(z)$ as
\begin{equation}M(z)=m_0\, \exp\left\{-\frac{\lambda}{c}\int_{z}^{\infty}U(s)\text{d}s\right\}\label{semiexplicitsoln},\end{equation}
which gives
\begin{equation} M(z)\to0 \quad \text{as} \quad z\to-\infty, \label{exBCMz} \end{equation}
and $M\leq m_0$ for all $z\in\mathbb{R}$. 
{Under the boundary condition $U(z)\to0$ as $z \to \infty$, at the leading edge of the travelling front (i.e. for $z \in (\ell, \infty)$ with $1 \ll \ell<\infty$ sufficiently large), we can use the ansatz
\begin{equation}
\label{Uansatz}
U(z) \approx \exp \left\{-\alpha \, z \right\},
\end{equation}
with $0 < \alpha < \infty$ for $z \in (\ell, \infty)$.
Inserting Equation~\eqref{Uansatz} into Equation~\eqref{semiexplicitsoln} we find
\begin{equation}
\label{M_SE_Uansatz}
M(z) \approx m_0 \exp \left\{-\dfrac{1}{\alpha} \, \left(\dfrac{\lambda \, U(z)}{c}\right)\right\},
\end{equation}
for $z \in (\ell, \infty)$.} Moreover, writing $\mathrm{d} U/\mathrm{d} z=V,$ we can rewrite Equations~\eqref{stdTW_u}-\eqref{stdTW_m} as a system of three first-order ordinary differential equations
\begin{align}{}
    \frac{\mathrm{d} U}{\mathrm{d} z}&=V,  \label{stdTW3_u}\\
    \frac{\mathrm{d} V}{\mathrm{d} z}&=\frac{1}{(1-M)}\bigg[-cV-\frac{\lambda}{c}M U V-\frac{\lambda^2}{c^2}MU^3 -U(1-U-M)\bigg],  \label{stdTW3_v}\\
    \frac{\mathrm{d} M}{\mathrm{d} z}&=\frac{\lambda}{c} M U, \label{stdTW3_m}
\end{align}
with boundary conditions given by
\begin{align}
    U(z)\to1, \quad V(z)\to 0 \quad &\text{and} \quad M(z)\to0 \quad \text{as} \quad z\to-\infty, \label{BCz3_1}\\
    U(z)\to0, \quad V(z)\to 0 \quad &\text{and} \quad  M(z)\to m_0 \quad \text{as} \quad z\to\infty. \label{BCz3_2}
\end{align}
The steady states of the system~\eqref{stdTW3_u}-\eqref{stdTW3_m} with boundary conditions~\eqref{BCz3_1}-\eqref{BCz3_2} are given by $\mathcal{S}_1=(1,0,0)$ and $\mathcal{S}_2=(0,0,m_0)$. Travelling wave analysis based on standard linear stability techniques (i.e. standard travelling wave analysis) \cite{curtin2020speed} seeks trajectories in the phase space that connect $\mathcal{S}_1$ at $z=-\infty$ to $\mathcal{S}_2$ at $z=\infty$~\cite{murray2001mathematical, lam2022introduction}.
The eigenvalues of the linearised system at $(U,V,M)=(1,0,0)$  are
\begin{equation}
    \sigma_1=\frac{\lambda}{c}, \quad \sigma_{2,3}=\frac{-c\pm\sqrt{c^2+4}}{2},
\end{equation}
which implies that $(1,0,0)$ is a {three-dimensional, hyperbolic, unstable saddle point \cite{anton2001calculus}, which has eigenvectors given by
\begin{align}
\mathbf{v_1}&=\begin{pmatrix} \dfrac{c^2-\lambda^2}{c^2(\lambda-1)+\lambda^2}, & \dfrac{\lambda(c^2-\lambda^2)}{c(c^2(\lambda-1)+\lambda^2)}, & 1 \end{pmatrix}^{T}  , \\
\mathbf{v_{2,3}}&=\begin{pmatrix} \dfrac{c\pm\sqrt{c^2+4}}{2}, & 0, & 1 \end{pmatrix}^{T} . \label{v2}
\end{align}}
The eigenvalues of the linearised system at  $(U,V,M)=(0,0,m_0)$ are
\begin{equation}\sigma_1=0, \quad \sigma_{2,3}=\frac{-c\pm\sqrt{c^2-4(1-m_0)^2}}{2(1-m_0)},\end{equation}
{with corresponding eigenvectors
\begin{align}
\mathbf{w_1}&=\begin{pmatrix} 0, & 0, & 1 \end{pmatrix}^{T}  , \\
\mathbf{w_{2,3}}&=\begin{pmatrix} \dfrac{c(c\pm\sqrt{c^2-4(1-m_0)^2})}{2\lambda m_0(m_0-1)}, & \dfrac{c(c^2\pm c\sqrt{c^2-4(1-m_0)^2}-2(1-m_0)^2)}{2\lambda m_0 (1-m_0)^2}, & 1 \end{pmatrix} ^{T} , \label{w3}
\end{align} 
which implies that $(0,0,m_0)$ is a stable, non-hyperbolic fixed point \cite{wiggins2003introduction} (see Appendix~\ref{APPeigs} for the full derivation). 
In all cases, we use the index $2$ to refer to the positive of the two choices, and $3$ for the negative.
When $c^2-4(1-m_0)^2\le0$, the steady state $(0,0,m_0)$ is a stable spiral as the eigenvalues have non-zero imaginary parts; however, when $c^2-4(1-m_0)^2\geq0$, the steady state is a stable node. 
In the case that the state $(0, 0,m_0)$ is a stable spiral, $U$ oscillates around this point on its approach and can therefore take negative values, see Figure~\ref{fig:cODEpp}. 
However, when $(0,0,m_0)$ is a stable node, there must exist a trajectory from $(1,0,0)$ to $(0,0,m_0)$ contained entirely in the region of phase space defined by $U\geq0$, $V\leq0$ and $M\geq0$, which ensures non-negativity of $U$ and $M$, as required to be biologically consistent. 
This demonstrates the existence of a minimum wave speed, $c_{\text{min}}=2(1-m_0),$ such that the dependent variables, $U$ and $M$, remain non-negative for all time.}
It is important to note that $c_{\text{min}}$ is a lower bound on the travelling wave speed, which is only actually attained for this system when the rescaled ECM degradation rate is sufficiently small, that is, $\lambda\to0^{+}$ (see Section~\ref{lam0}). 
This is clearly shown in Figure~\ref{fig:cvsdmvsm0}, which also demonstrates that decreasing $m_0$ results in an increase in the travelling wave speed.

\begin{figure}[h!]
    \centering
    \includegraphics[scale=0.55]{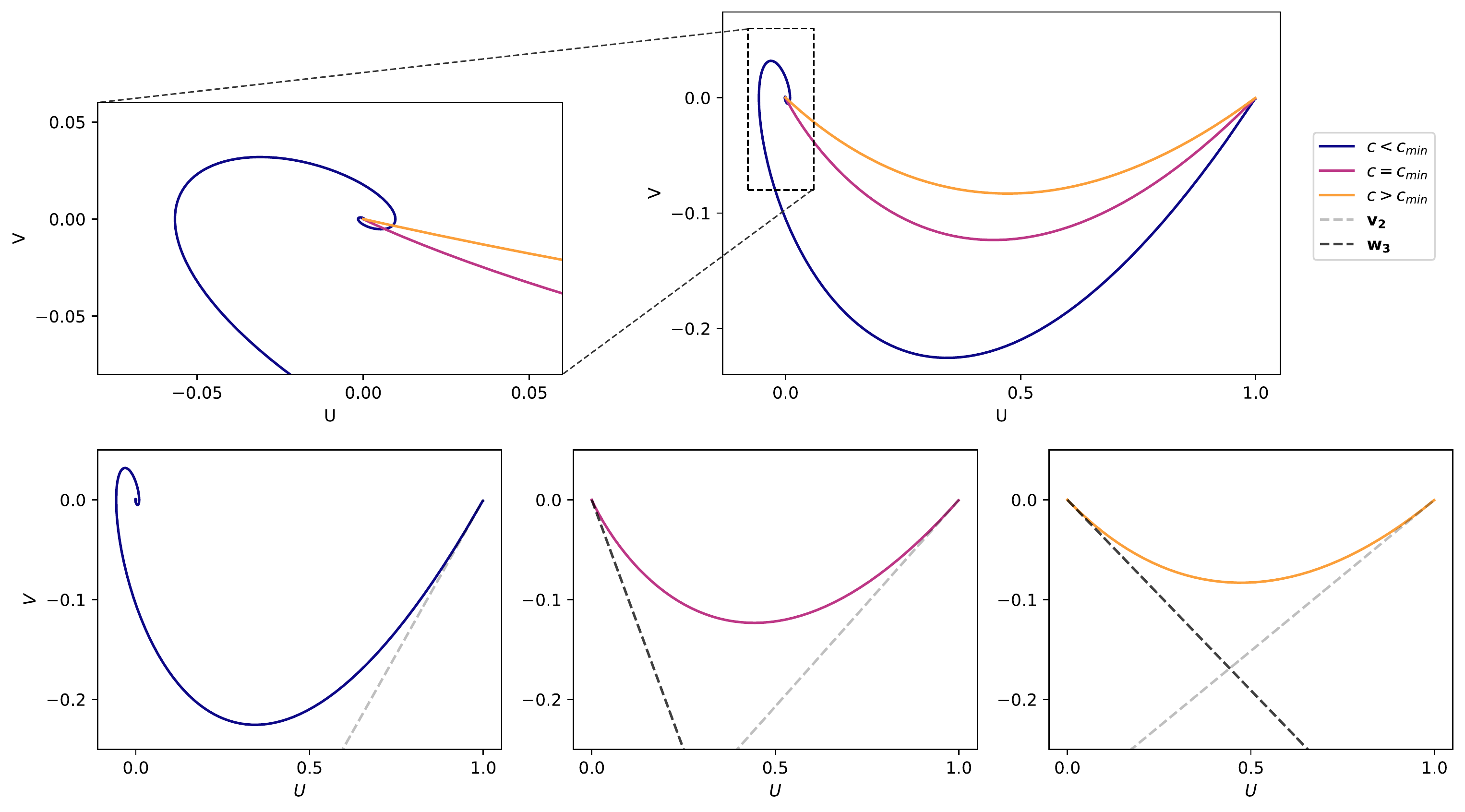}
    \caption{Phase plane plot of the ODE system~\eqref{stdTW3_u}-\eqref{stdTW3_v}, for different travelling wave speeds, $c$, demonstrating the change from a stable spiral to a stable node as the travelling wave speed exceeds $c_{\text{min}}.$ The corresponding unstable eigenvector given by Equation~\eqref{v2} and stable eigenvector given by Equation~\eqref{w3} are overlaid in the lower plots~\cite{curtin2020speed}. Further specifics of the parameter values and the numerical methods used can be found in Appendix~\ref{appNS}.}
    \label{fig:cODEpp}
\end{figure}

Travelling wave analysis has also been performed on a PDE model for melanoma invasion into the skin \cite{browning2019bayesian}, where volume-filling effects of cells are not considered to impact cell movement \cite{el2021travelling}, as described earlier. 
Since travelling wave analysis is always performed on the linearised system, it follows that, the additional term describing cell movement prevented by the presence of other cells is lost from Equation~\eqref{stdTW3_v} during linearisation and the minimum travelling wave speed is the same as that derived in \cite{el2021travelling}: $c_{\text{min}}=2(1-m_0)$. 
Another minimal model for tumour growth was proposed in \cite{colson_travelling-wave_2021}, where the volume-filling effects of cells were not accounted for in describing cell movement or cell proliferation. 
Both of these models have the same equation for ECM density as Equation~\eqref{NDeqn2_m}, and the models in \cite{ el2021travelling, colson_travelling-wave_2021} have the same flux terms in the equation for cell density evolution, but the model in \cite{colson_travelling-wave_2021} has one less reaction term since proliferation is unimpeded by the local ECM density.
As a result of the fact that all volume-filling effects are encoded in non-linear terms, changes to the flux terms alone (within this suite of models) have no effect on the predicted minimum travelling wave speed, as they are all identical after linearisation. 
However, alterations to the net proliferation terms do significantly impact the minimum travelling wave speeds predicted by standard travelling wave analysis. Further information regarding these models and their differences can be found in Appendix~\ref{APPcomp}.

As previously described, we are particularly interested in investigating the dependence of travelling wave solutions on the parameters $\lambda$, the rescaled ECM degradation rate, and $m_0$, the density of ECM far ahead of the wave. 
Having now determined that the minimum travelling wave speed decreases linearly as $m_0$ increases, we now aim to explore the relationship between the numerically estimated travelling wave speed and $\lambda$. 

Since the travelling wave speed depends on $\lambda$,  standard perturbation techniques are difficult to apply to the travelling wave Equations~\eqref{stdTW_u}-\eqref{stdTW_m}. 
As a result, we examine Figure~\ref{fig:cvsdmvsm0} for clues as to how to proceed. 
We immediately see that for sufficiently small $\lambda$ it appears that the numerically estimated travelling wave speed is independent of $\lambda$ and matches the speed predicted by standard travelling wave analysis. 
It can also be seen from the contour plot in Figure~\ref{fig:cvsdmvsm0} that for large values of $\lambda$, the speed converges for all values of $m_0\in[0,1)$. 
As such, we now investigate the asymptotic limits corresponding to slow and fast rescaled ECM degradation rates,  $\lambda\to0^{+}$ and $\lambda\to\infty$, respectively.

\begin{subsection}{{Formal a}symptotic analysis for $\lambda\to0^{+}$} \label{lam0} 
{Using Equation~\eqref{M_SE_Uansatz} it is clear that
\begin{equation}
M(z)\approx m_0 \exp \left\{-\dfrac{1}{\alpha} \, \left(\dfrac{\lambda \, U(z)}{c}\right)\right\}  \to m_0 \quad \text{as} \quad \lambda\to0^{+}, \label{Mzm0}
\end{equation}
for $z \in (\ell, \infty)$} (see Figure~\ref{fig:small_lam} {or Figures~\ref{fig:small_lam_TW} and~\ref{fig:zoomed_small_lam} for the travelling wave profiles}).

\begin{figure}[h!]
    \centering
    \includegraphics[scale=0.575]{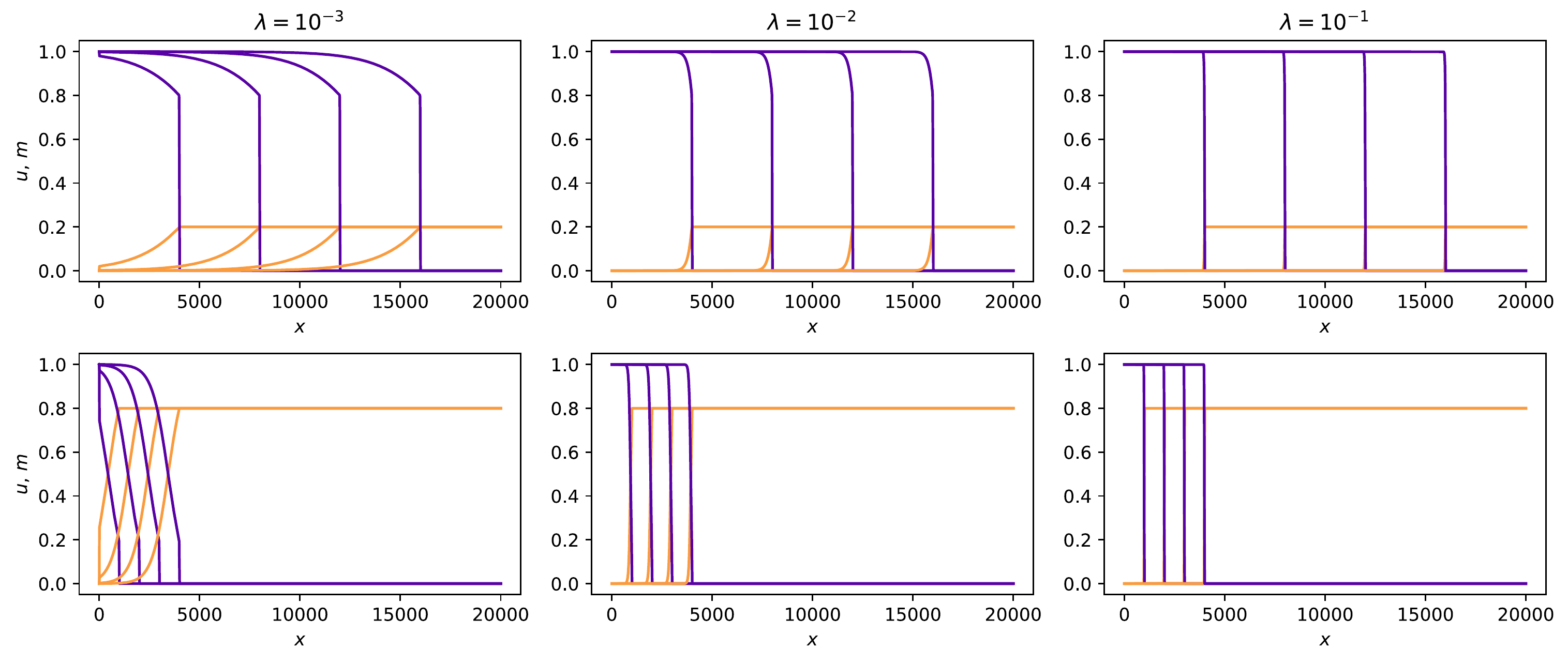}
    \caption{{Numerical solutions of Equations~\eqref{NDeqn2_u}-\eqref{NDeqn2_m} subject to the initial conditions~\eqref{hIC_u}-\eqref{hIC_m}, for $m_0=0.2$ in the top row and $m_0=0.8$ in the bottom row, and for rescaled ECM degradation rates $\lambda=10^{-3}, \,10^{-2},\,10^{-1}.$ Cell densities are shown in purple and ECM densities in orange at times $t=2500,\,5000,\,7500,\,10000$ from left to right. Further specifics of the parameter values and the numerical methods used can be found in Appendix~\ref{appNS}.}}
    \label{fig:small_lam}
\end{figure}

{In the asymptotic regime $\lambda \to 0^+$, substituting Equation~\eqref{Mzm0} into Equation~\eqref{subst_TW} and using the fact that, since $0\leq U(z)<1$ for $z \in (\ell, \infty)$ and ${{\rm d} U(z)}/{{\rm d} z} \approx - \alpha \, U(z)$ for $z \in (\ell, \infty)$ (cf. the ansatz given by Equation~\eqref{Uansatz}), the following asymptotic relation holds
\begin{equation}
m_0 \exp \left\{-\dfrac{1}{\alpha} \, \left(\dfrac{\lambda \, U(z)}{c}\right)\right\} \left[U(z) \left(\dfrac{\lambda \, U(z)}{c} \right)^2 + \dfrac{{\rm d} U(z)}{{\rm d}z} \left(\dfrac{\lambda \, U(z)}{c} \right)\right] \to 0 \quad \text{as} \quad \lambda\to0^{+},
\end{equation}
for $z \in (\ell, \infty)$, we find
\begin{equation}
\left(1 - m_0 \right) \dfrac{{\rm d}^2 U(z)}{{\rm d}z^2} + c \dfrac{{\rm d} U(z)}{{\rm d}z} + U(z) \big[\left(1 - m_0\right) - U(z) \big] \approx 0, \label{rescaledFKPP}
\end{equation}
for $z \in (\ell, \infty)$.
}
Equation~\eqref{rescaledFKPP} is equivalent to the FKPP model~\eqref{FKPP} in travelling wave co-ordinates
\begin{equation}\label{rescaledFKPPog}
    \hat{D}\frac{\mathrm{d}^2\hat{U}}{\mathrm{d} z}+\hat{c}\frac{\mathrm{d}\hat{U}}{\mathrm{d}z}+\hat{r}\hat{U}\bigg(1-\frac{\hat{U}}{\hat{K}}\bigg)=0,
\end{equation}
with $\hat{D}=\hat{r}=\hat{K}=1-m_0,$ with $\hat{c}_{\text{min}} =2(1-m_0),$ as predicted earlier. 
An excellent match {around the leading edge of the travelling wave front} between the travelling wave solution to the FKPP model~\eqref{FKPP} {{(Equation~\eqref{rescaledFKPPog} in travelling wave co-ordinates)}} and {Equations}~\eqref{NDeqn2_u}-\eqref{NDeqn2_m} for low values of the rescaled ECM degradation rate can be seen in the plot on the left in Figure~\ref{fig:increasingdmvsfkpp} {- see also Figure~\ref{fig:small_lam} or Figures~\ref{fig:small_lam_TW} and~\ref{fig:zoomed_small_lam}}. 

{We now consider the region $z\in(-\infty, \ell)$ by rescaling Equations~\eqref{stdTW_u}-\eqref{stdTW_m} using the new variable $\epsilon=z\lambda$ for $\epsilon\in(-\infty, \ell\lambda]$. The system of Equations~\eqref{stdTW_u}-\eqref{stdTW_m} becomes
\begin{align}
    -c\lambda\frac{\mathrm{d}U}{\mathrm{d}\epsilon}&=\lambda^2\frac{\mathrm{d}}{\mathrm{d}\epsilon}\bigg[(1-M)\frac{\mathrm{d}U}{\mathrm{d}\epsilon}+U\frac{\mathrm{d}M}{\mathrm{d}\epsilon}\bigg]+U(1-U-M), \label{TW_eps_u} \\
    \lambda\frac{\mathrm{d}M}{\mathrm{d}\epsilon}&=\frac{\lambda}{c}MU. \label{TW_eps_m}
\end{align}
For $\lambda\to0^{+}$, we find from Equation~\eqref{TW_eps_u} that $U(\epsilon)(1-U(\epsilon)-M(\epsilon))=0$, so that for $\epsilon\in(-\infty, \ell\lambda]$ we have $U(\epsilon)=1-M(\epsilon)$ since $U(\epsilon)\to1$ as $\epsilon\to-\infty$.
By substitution into Equation~\eqref{TW_eps_m}, we find 
$$\frac{\mathrm{d}M}{\mathrm{d}\epsilon}=\frac{M(1-M)}{c},$$ which, using the matching condition that $M(\epsilon=\ell\lambda)=m_0$, gives 
\begin{equation}
    M(\epsilon)=\dfrac{m_0 \,\text{exp}\{-(\lambda\ell-\epsilon)/c\}}{1-m_0+m_0\,\text{exp}\{-(\lambda\ell-\epsilon)/c\}}, \label{M_eps}
\end{equation}
Recalling that $U(\epsilon)=1-M(\epsilon)$, we obtain
\begin{equation}
    U(\epsilon)=\dfrac{1-m_0}{1-m_0+m_0\,\text{exp}\{-(\lambda\ell-\epsilon)/c\}}, \label{U_eps}
\end{equation}
which tends to $1$ as $\epsilon\to-\infty$ and to $1-m_0$ as $\epsilon\to\ell\lambda.$
In the travelling wave co-ordinate, $z$, Equation~\eqref{U_eps} can be written as
\begin{equation}
 U(z)=\dfrac{1-m_0}{1-m_0+m_0\, \text{exp}\{-{\lambda}(\ell-z)/c\}}, \label{anau}
\end{equation}
for $z\in(-\infty, \ell]$ and the solution to the FKPP model, as given by Equation~\eqref{rescaledFKPP}, for $z\in(\ell, \infty)$. 
In the travelling wave co-ordinate, $z$, the solution for the wave profile of the ECM given by Equation~\eqref{M_eps} is
\begin{equation}
M(z)=\dfrac{m_0\,\text{exp}\{-{\lambda}(\ell-z)/c\}}{1-m_0+m_0\,\text{exp}\{-{\lambda}(\ell-z)/c\}}, \label{anam}
\end{equation}
for $z\in(-\infty, \ell],$ and $M(z)=m_0$ for $z\in(\ell, \infty)$, as given by Equation~\eqref{Mzm0}.
An excellent agreement between these analytical solutions and the numerical results can be observed in Figure~\ref{fig:small_lam_match_ana}.
}

\begin{figure}[h!]
    \centering    
    \includegraphics[scale=0.5]{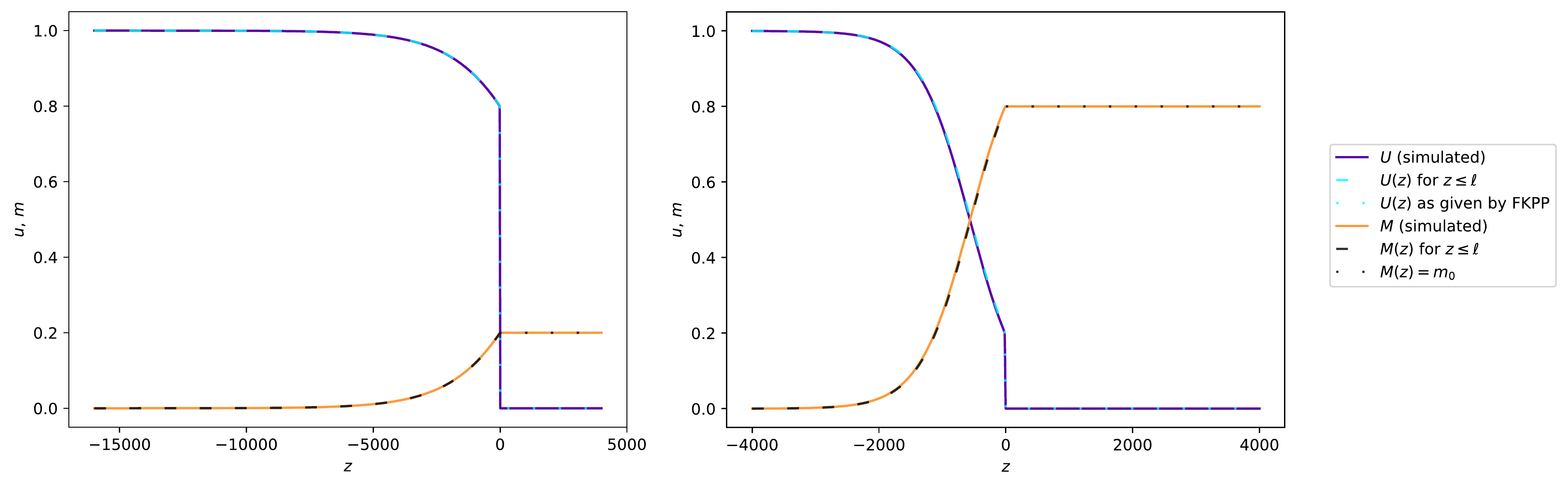}
    \caption{{Numerical solutions of Equations~\eqref{NDeqn2_u}-\eqref{NDeqn2_m} subject to the initial conditions~\eqref{hIC_u}-\eqref{hIC_m}, for $m_0=0.2$ on the left and $m_0=0.8$ on the right with rescaled ECM degradation rate $\lambda=10^{-3}$ translated into the travelling wave co-ordinate, $z$. {Solid lines represent the cell and ECM densities from numerical simulations in purple and orange respectively. The FKPP solution~\eqref{rescaledFKPP} in travelling wave co-ordinates is plotted as a dotted blue line. The solution $M(z)=m_0$ is plotted in dotted black, and the analytical solutions given by Equations~\eqref{anau} and~\eqref{anam} are plotted in dashed blue and black lines, respectively. Further specifics of the parameter values and the numerical methods used can be found in Appendix~\ref{appNS}.}}}
    \label{fig:small_lam_match_ana}
\end{figure}

Similar models, such as those described at the end of Section~\ref{TWA} that do not have volume-filling effects taken into account, demonstrate qualitatively similar behaviour. 
In all of these models, at very low rescaled ECM degradation rates we observe convergence of the solutions to those of the FKPP model with rescaled parameters. 
For models with the same cell proliferation term as in Equation~\eqref{NDeqn2_u}, the rescaled parameters are the same and the convergence has qualitatively similar behaviour, as displayed in the plot on the left in Figure~\ref{fig:increasingdmvsfkpp}. 
As a result, in the limit of very small rescaled ECM degradation rates, $\lambda\to0^{+}$, the model~\eqref{NDeqn2_u}-\eqref{NDeqn2_m} can be simplified to that presented in \cite{el2021travelling}, which neglects the volume-filling effects of cells upon cell movement. 
This model can, in turn, be well approximated by the FKPP model~\eqref{FKPP} with rescaled parameters $\hat{D}=\hat{r}=\hat{K}=1-m_0.$ 
This result is consistent with predictions from standard travelling wave analysis. 
However, for the model presented in \cite{colson_travelling-wave_2021}, the parameters of the rescaled FKPP model to which the model converges are, instead, $\hat{D}=1-m_0$ and $\hat{r}=\hat{K}=1,$ that entails a higher cell carrying capacity density since proliferation is not impacted by the surrounding ECM. 
See Appendix~\ref{APPcomp} for a more detailed comparison.
As such, the model~\eqref{NDeqn2_u}-\eqref{NDeqn2_m} is poorly approximated using models,  such as that in \cite{colson_travelling-wave_2021}, with different underlying assumptions for cell proliferation. 
These differences highlight the importance of fully laying out all of the model assumptions at the single-cell level before deriving the PDE model, so that the population-level model fully captures behaviours associated with the underlying cell-level assumptions, in all parameter regimes. 
\begin{figure}[h!]
    \centering
\hspace*{-.3cm} 
    \includegraphics[scale=0.52]{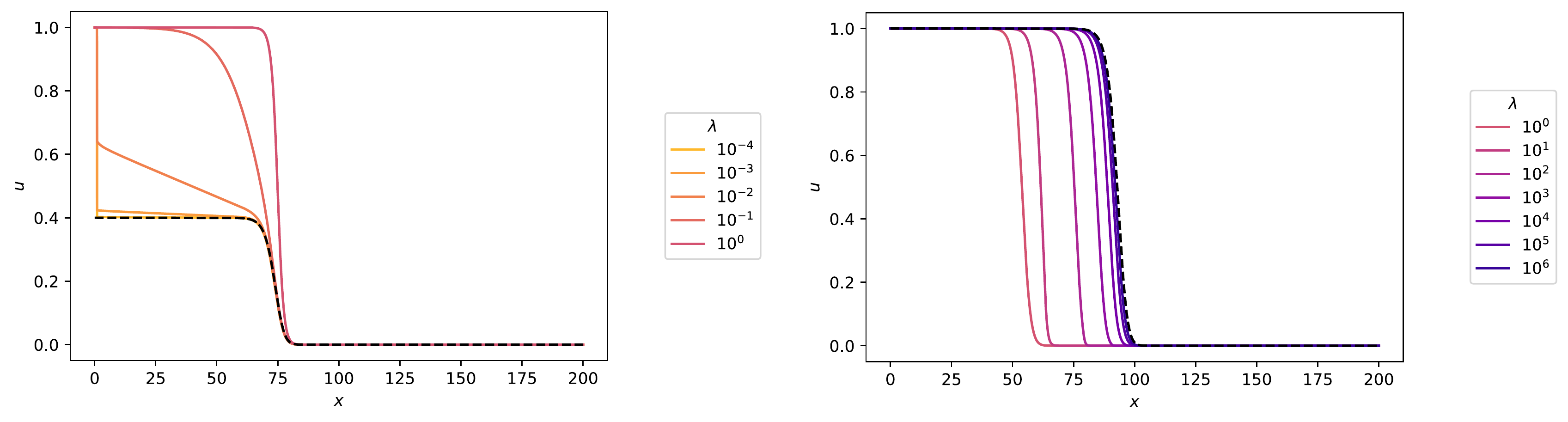}
    \caption{Left: plot of the cell density, $u$, obtained through numerical simulations of Equations~\eqref{NDeqn2_u}-\eqref{NDeqn2_m} subject to the initial conditions~\eqref{hIC_u}-\eqref{hIC_m} (solid lines) for small values of $\lambda$, and numerical simulations of the FKPP model~\eqref{FKPP} with rescaled coefficients $\hat{D}=\hat{r}=\hat{K}=1-m_0$ (dashed black line) with $t=100$ and $m_0=0.6$. Right: plot of the cell density, $u$, obtained through numerical simulations of Equations~\eqref{NDeqn2_u}-\eqref{NDeqn2_m} subject to the initial conditions~\eqref{hIC_u}-\eqref{hIC_m} (solid lines) for large values of $\lambda$, and numerical simulations of the FKPP model~\eqref{FKPP1} (dashed black line) in the plot on the right for $t=50$ and $m_0=0.4$. Qualitatively, the same behaviour is observed for all $m_0\in[0,1)$. Further specifics of the parameter values and the numerical methods used for the simulations can be found in Appendix~\ref{appNS}.}
    \label{fig:increasingdmvsfkpp}
\end{figure}

\end{subsection}

\begin{subsection}{Formal asymptotic analysis for $\lambda\to\infty$} \label{laminf}
In the case of very large rates of ECM degradation, by considering the semi-explicit solution for $M$ in terms of $U$ given by Equation~\eqref{M_SE_Uansatz}, 
{we see that
\begin{equation}
M(z)\approx m_0 \exp \left\{-\dfrac{1}{\alpha} \, \left(\dfrac{\lambda \, U(z)}{c}\right)\right\} \to 0 \quad \text{as} \quad \lambda\to\infty, \label{Mlarge}
\end{equation}
 for $z \in (\ell, \infty)$ {(see Figure~\ref{fig:large_lam} or Figure~\ref{fig:large_lam_TW} for the travelling wave profiles)}.
In the asymptotic regime $\lambda\to\infty$, substituting Equation~\eqref{Mlarge} into Equation~\eqref{subst_TW} and using the fact that, since $0\leq U(z)<1$ for $z \in (\ell, \infty)$ and ${{\rm d} U(z)}/{{\rm d} z} \approx - \alpha \, U(z)$ for $z \in (\ell, \infty)$ (cf. the ansatz given by Equation~\eqref{Uansatz}), the following asymptotic relation holds
\begin{equation}
m_0 \exp \left\{-\dfrac{1}{\alpha} \, \left(\dfrac{\lambda \, U(z)}{c}\right)\right\} \left[U(z) \left(\dfrac{\lambda \, U(z)}{c} \right)^2 + \dfrac{{\rm d} U(z)}{{\rm d}z} \left(\dfrac{\lambda \, U(z)}{c} \right)\right] \to 0 \quad \text{as} \quad \lambda\to\infty,
\end{equation}
for $z \in (\ell, \infty)$, we find  
\begin{equation}
\dfrac{{\rm d}^2 U(z)}{{\rm d}z^2} + c \dfrac{{\rm d} U(z)}{{\rm d}z} + U(z) \big(1 - U(z) \big) \approx 0, 
\end{equation}
for $z \in (\ell, \infty)$.
Hence, when $\lambda \to \infty$ we expect $U(z)$ at the leading edge of the travelling front to behave, to a first approximation, as the solution to the FKPP Equation~\eqref{FKPP1}} in travelling wave co-ordinates {subject to the boundary condition~\eqref{BCU2z}, for which $c_{\text{min}}=2.$ 
This result can also be observed numerically in the plot on the right of Figure~\ref{fig:increasingdmvsfkpp}.}  The same behaviour is observed in similar models without volume-filling effects \cite{el2021travelling, colson_travelling-wave_2021}, demonstrating that the model~\eqref{NDeqn2_u}-\eqref{NDeqn2_m} can be approximated, to an extent, with any of these simpler models in the parameter regime $\lambda\to\infty,$ {as growth and diffusion are unrestricted by the ECM within a neighbourhood of the travelling wave front.}
\begin{figure}[h!]
    \centering
    \includegraphics[scale=0.575]{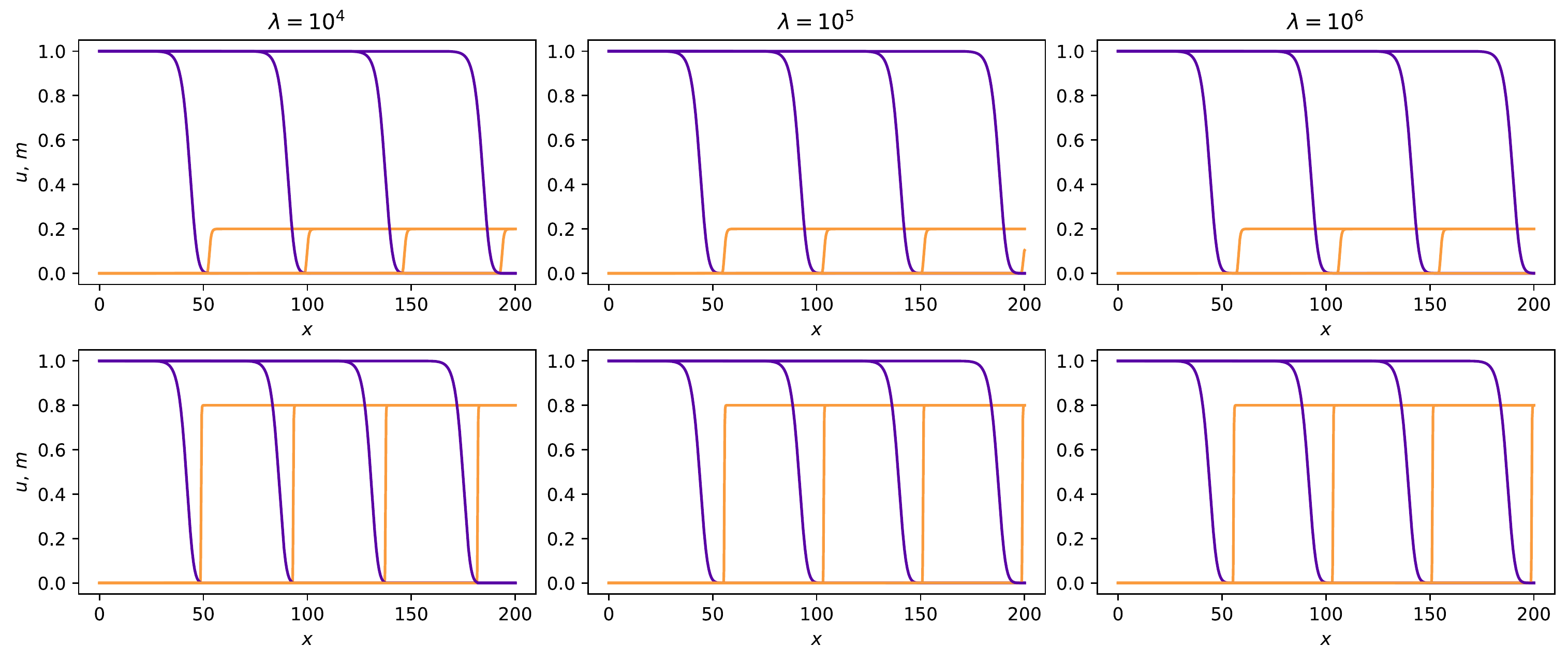}
    \caption{{Numerical solutions of Equations~\eqref{NDeqn2_u}-\eqref{NDeqn2_m} subject to the initial conditions~\eqref{hIC_u}-\eqref{hIC_m}, for $m_0=0.2$ in the top row and $m_0=0.8$ in the bottom row, and for rescaled ECM degradation rates $\lambda=10^{4}, \,10^{5},\,10^{6}.$ Cell densities are shown in purple and ECM densities in orange at times $t=25,\,50,\,75,\,100$ from left to right. Further specifics of the parameter values and the numerical methods used can be found in Appendix~\ref{appNS}.}}
    \label{fig:large_lam}
\end{figure}

\end{subsection}
\end{section}

\begin{section}{Discussion and conclusions}
In this paper, a model for cell invasion into the surrounding ECM has been studied by considering primarily its travelling wave solutions. In this model, derived from first principles from an agent-based model describing cell-level behaviours, cells evolve under the action of diffusion and proliferation, that is coupled to degradation of the surrounding ECM. As a result of volume-fillling effects, cells require space ahead of the wave front in order to invade the domain.

Numerical solutions of the PDE model~\eqref{NDeqn2_u}-\eqref{NDeqn2_m} demonstrate a complex relationship between the travelling wave speed, $c$, the density of ECM far ahead of the wave of cells, $m_0$, and the rescaled ECM degradation rate, $\lambda.$ Partial relationships between these parameters in asymptotic regimes of interest have been established, including that $c\to2(1-m_0)$ as $\lambda\to0^{+},$ and that $c\to2^{-}$ as $\lambda\to\infty$.
A good agreement with the FKPP model~\eqref{FKPP} has been demonstrated in the case where $\lambda\to\infty,$ and we showed that the impacts of introducing volume-filling effects of cells to reduce cell movement (in comparison to the model in \cite{el2021travelling}) are minimal. 
As such, the FKPP model~\eqref{FKPP} provides a suitable model simplification to reproduce the qualitative behaviours of the fully dimensional system in the case of a large ECM degradation rate, $\tilde{\lambda}$, compared to the proliferation rate, $\tilde{r}$. 
Since $\lambda={\tilde{\lambda}{\tilde{K}}}/{\tilde{r}}$, the results equivalently suggest that as $\tilde{K}\to\infty$, the system can be well modelled by the FKPP model~\eqref{FKPP}. 
This describes a model where volume-filling effects are negligible, and thus the speed of the invasion front is given by $c_{\text{min}}=2.$   
For $\lambda\to0^{+}$, which is representative of very large proliferation rates compared to the rescaled ECM degradation rates, or extremely small carrying capacities, the system can be studied by considering the simplification to a rescaled FKPP model~\eqref{rescaledFKPP}. In this case, travelling waves are observed for $m\in[0,1)$, but the speed of the invasion front is now given by $c_{\text{min}}=2(1-m_0).$
Converting back to dimensional variables, as with the FKPP model~\eqref{FKPP}, the analytically predicted travelling wave speed increases with the cell proliferation rate, but with a more complicated relationship for the regions of parameter space corresponding to where the relationship between the travelling wave speed and rescaled ECM degradation rate is not yet well established.
It is likely this complicated relationship indicates that the system exhibits changes between pulled, pushed and semi-pushed waves due to the non-linear cross-species dynamics that vary in strength for different parameter values \cite{birzu2018fluctuations}. This could be investigated further by examining the ratio between the travelling wave speeds for different parameter values.

It is also clear that qualitatively similar results are observed between this new model with volume-filling, and previously studied models outside this framework, as described by Table~\ref{table:modelcomp}, in all cases where $m_0\in[0,1)$. 
Therefore, it could be said that the model originally proposed in \cite{browning2019bayesian} provides a good model simplification for any case where $m_0\in[0,1)$. 
In the case where $m_0=1$, the region that is initially uninvaded by cells is full with ECM, such that proliferation and movement of cells into this region is entirely prevented. 
This result provides the starkest difference between the model studied in this paper and those previously studied elsewhere \cite{el2021travelling, colson_travelling-wave_2021}. 
It is observed that in the case of compactly-supported initial cell density, cell invasion cannot occur into the region where $m(x,0)=1,\, u(x,0)=0,$ and thus pinning occurs and travelling waves cannot form \cite{wang2019pinned}.
It is biologically reasonable to assume that an invading cell population might have zero density far ahead of the invading front.
However it is important to note that the model considered here is a very simplistic model for cell invasion into ECM, and if further biological complications, such as the secretion of matrix metalloproteinases (MMPs) by cells to degrade and remodel ECM, were introduced then these phenomenological results would no longer be observed \cite{perumpanani_extracellular_1998}. 
This is because we could reasonably assume MMPs could still diffuse into regions occupied entirely by ECM, and then degrade it. 

\begin{table}[h!]
\begin{center}
\begin{tabular}{ |c||c|c|c|c|c|c| } 
\hline
\rule{0pt}{10pt} \multirow{3}{*}{\textbf{Model}} & \multicolumn{2}{c|}{Volume-filling}  & \multirow{3}{*}{Diffusion term} & \multicolumn{2}{c|}{Volume-filling} & \multirow{3}{*}{Reaction term}   \\
& \multicolumn{2}{c|}{in movement} & & \multicolumn{2}{c|}{in proliferation} &  \\
& by cells & by ECM  & & by cells & by ECM &\\
\hline 
\rule{0pt}{20pt} Colson \cite{colson_travelling-wave_2021} & - & + & $\frac{\partial}{\partial x}\bigg[ (1- m)\frac{\partial u}{\partial x}\bigg]$ & + & - &  $u(1-u)$\\ 
\rule{0pt}{20pt} Browning \cite{el2021travelling, browning2019bayesian} & - & + & $\frac{\partial}{\partial x}\bigg[ (1- m)\frac{\partial u}{\partial x}$\bigg] & + & + & $u(1-u-m)$ \\ 
\rule{0pt}{20pt} Equations~\eqref{NDeqn2_u}-\eqref{NDeqn2_m} & + & + & $\frac{\partial}{\partial x}\bigg[ (1- m)\frac{\partial u}{\partial x}+u\frac{\partial m}{\partial x}\bigg]$ & +  & +  & $u(1-u-m)$ \\ 
 \hline
\end{tabular}
\end{center}
\caption{Description of the volume-filling effects of cells and ECM considered by the models compared in this study.}
\label{table:modelcomp}
\end{table}

The overall conclusion of our study is that there exist simpler models for cell invasion into ECM such as \cite{fisher_wave_1937, el2021travelling, colson_travelling-wave_2021}, that are defined by similar guiding principles and can be used to reproduce the qualitative behaviours of the travelling waves observed in the model presented in this work. 
Analysis of these systems confirms that the qualitative model predictions are conserved, and therefore the simpler models can be used in future studies to reduce computational complexity and make the resulting PDE model more analytically tractable.
The disadvantage of this conclusion, however, is that in order to use these models to infer parameters from data, extra steps would be required to validate whether the correct model has been selected. For example, analysis of cell trajectories can help infer the cell-cell interactions underlying the motility mechanism, and distinguish between the suite of models with qualitatively similar behaviours \cite{simpson2009pathlines, bowden2013design, ross2015inference}. 
Our results reveal that the reaction term significantly impacts the travelling wave speed for small and intermediate values of $\lambda$ and thus, it could be used to inform model development, by defining the reaction term by considering whether space or nutrients are the limiting factor for cell invasion into ECM; and model selection, by comparing the expected wave speeds to the data. 

There are a variety of possible extensions to the work presented in this paper. The underlying on-lattice agent-based model of cell movement involves a number of simplifying assumptions, such as that cells can only degrade ECM agents in the same lattice site. 
By varying these assumptions, there would be the possibility to expand the biological applicability of the study to determine under which regimes the resulting models can also be approximated by simpler seminal models of cell invasion.
Different proliferation terms, as well as terms to account for ECM evolution in more detail could be included, such as ECM remodelling by cells, or elastic deformation \cite{malik2020impact}.
Beyond this, another clear extension of this work would be to introduce further spatial dimensions, or different geometries, that are particularly interesting for studying cancer cell invasion, and to investigate the stability of the travelling wave solutions for the different possible models. 
For the case $\lambda\to0^{+}$, there is an opportunity to apply boundary layer theory and asymptotic analysis to arrive at an expression for the full travelling wave profile at long times.
It would also be of particular interest to arrive at some functional form for the travelling wave speed, $c(\lambda, m_0)$, for all possible parameter values, and to define the critical value of $\lambda_c$, depending on $m_0$ (see Figure \ref{fig:cvsdmvsm0}), whereby for $\lambda<\lambda_c$ the minimum travelling wave speed observed numerically matches that predicted by standard travelling wave analysis $c=c_{\text{min}}$. 
The critical value, $\lambda_c$, might be found by establishing the basins of attraction for each steady state and seeking parameter regimes where the dynamics follow different paths.  
If possible, this knowledge could then further aid an investigation using perturbation methods into the shape of the wave front for intermediate values of $\lambda$ and by characterising this behaviour, this model could be used to describe biological scenarios such as tumour growth, where $\lambda$ would represent the rate at which the tumour cells were able to degrade ECM in the surrounding envrionment. 

\end{section}

\section*{Acknowledgements}
R. M. C. is supported by funding from the Engineering and Physical Sciences Research Council (EPSRC) and Wolfson College, University of Oxford.
T. L. gratefully acknowledges support from the Italian Ministry of University and Research (MUR) through the grant ``Dipartimenti di Eccellenza 2018-2022'' (Project no. E11G18000350001), the PRIN 2020 project (No. 2020JLWP23) ``Integrated Mathematical Approaches to Socio–Epidemiological Dynamics'' (CUP: E15F21005420006), and the INdAM group GNFM.
The authors are grateful to Kevin Painter and Chloe Colson for interesting discussions regarding travelling waves in cell invasion models. 

\printbibliography

@article{giniunaite2020modelling,
  title={Modelling collective cell migration: Neural crest as a model paradigm},
  author={Gini{\=u}nait{\.e}, R. and Baker, R. E. and Kulesa, P. M. and Maini, P. K.},
  journal={Journal of Mathematical Biology},
  volume={80},
  pages={481--504},
  year={2020},
  publisher={Springer}
}

@article{klm2015hybrid,
  title={Hybrid models of cell and tissue dynamics in tumor growth},
  author={Kim, Y. and Othmer, H. G.},
  journal={Mathematical Biosciences and Engineering},
  volume={12},
  number={6},
  pages={1141},
  year={2015},
  publisher={NIH Public Access}
}

@article{wang2019pinned,
  title={Pinned, locked, pushed, and pulled traveling waves in structured environments},
  author={Wang, C.-H. and Matin, S. and George, A. B. and Korolev, K. S.},
  journal={Theoretical Population Biology},
  volume={127},
  pages={102--119},
  year={2019},
  publisher={Elsevier}
}

@article{birzu2018fluctuations,
  title={Fluctuations uncover a distinct class of traveling waves},
  author={Birzu, G. and Hallatschek, O. and Korolev, K. S.},
  journal={Proceedings of the National Academy of Sciences},
  volume={115},
  number={16},
  pages={E3645--E3654},
  year={2018},
  publisher={National Academy of Sciences}
}

@article{curtin2020speed,
  title={Speed switch in glioblastoma growth rate due to enhanced hypoxia-induced migration},
  author={Curtin, L. and Hawkins-Daarud, A. and Van Der Zee, K. G. and Swanson, K. R. and Owen, M. R.},
  journal={Bulletin of Mathematical Biology},
  volume={82},
  number={3},
  pages={43},
  year={2020},
  publisher={Springer}
}

@article{simpson2009pathlines,
  title={Pathlines in exclusion processes},
  author={Simpson, M. J. and Landman, K. A. and Hughes, B. D.},
  journal={Physical Review E},
  volume={79},
  number={3},
  pages={031920},
  year={2009},
  publisher={APS}
}

@article{bowden2013design,
  title={Design and interpretation of cell trajectory assays},
  author={Bowden, L. G. and Simpson, M. J. and Baker, R. E.},
  journal={Journal of the Royal Society Interface},
  volume={10},
  number={88},
  pages={20130630},
  year={2013},
  publisher={The Royal Society}
}

@article{ross2015inference,
  title={Inference of cell--cell interactions from population density characteristics and cell trajectories on static and growing domains},
  author={Ross, R. J. H. and Yates, C. A. and Baker, R. E.},
  journal={Mathematical Biosciences},
  volume={264},
  pages={108--118},
  year={2015},
  publisher={Elsevier}
}

@article{penington2011building,
	author = {Penington, C. J. and Hughes, B. D. and Landman, K. A.},
	journal = {Physical Review E},
	number = {4},
	pages = {041120},
	publisher = {APS},
	title = {Building macroscale models from microscale probabilistic models: A general probabilistic approach for nonlinear diffusion and multispecies phenomena},
	volume = {84},
	year = {2011}}

@article{sengers2007experimental,
	author = {Sengers, B. G. and Please, C. P. and Oreffo, R. O. C.},
	journal = {Journal of the Royal Society Interface},
	number = {17},
	pages = {1107--1117},
	publisher = {The Royal Society London},
	title = {Experimental characterization and computational modelling of two-dimensional cell spreading for skeletal regeneration},
	volume = {4},
	year = {2007}}

@article{martin2010tumour,
	author = {Martin, N. K. and Gaffney, E. A. and Gatenby, R. A. and Maini, P. K.},
	journal = {Journal of Theoretical Biology},
	number = {3},
	pages = {461--470},
	publisher = {Elsevier},
	title = {Tumour--stromal interactions in acid-mediated invasion: A mathematical model},
	volume = {267},
	year = {2010}}

@article{mcgillen2014general,
	author = {McGillen, J. B. and Gaffney, E. A. and Martin, N. K. and Maini, P. K.},
	journal = {Journal of Mathematical Biology},
	number = {5},
	pages = {1199--1224},
	publisher = {Springer},
	title = {A general reaction--diffusion model of acidity in cancer invasion},
	volume = {68},
	year = {2014}}

@article{gurtin1977diffusion,
	author = {Gurtin, M. E. and MacCamy, R. C.},
	journal = {Mathematical Biosciences},
	number = {1-2},
	pages = {35--49},
	publisher = {Elsevier},
	title = {On the diffusion of biological populations},
	volume = {33},
	year = {1977}}

@article{johnston2017co,
	author = {Johnston, S. T. and Baker, R. E. and McElwain, D. L. and Simpson, M. J.},
	journal = {Scientific Reports},
	number = {1},
	pages = {1--19},
	publisher = {Nature Publishing Group},
	title = {Co-operation, competition and crowding: a discrete framework linking Allee kinetics, nonlinear diffusion, shocks and sharp-fronted travelling waves},
	volume = {7},
	year = {2017}}

@book{anton2001calculus,
  title={Calculus: {M}ultivariable version},
  author={Anton, H. and Bivens, I. and Davis, S.},
  publisher = {Von Hoffmann Press},
  year={2001}
}

@article{morris2020identifying,
	author = {Morris, B. and Curtin, L. and Hawkins-Daarud, A. and Hubbard, M. E. and Rahman, R. and Smith, S. J. and Auer, D. and Tran, N. L. and Hu, L. S. and Eschbacher, J. M. and Smith, K. A. and Stokes, A. and Swanson, K. R. and Owen, M. R.},
	journal = {Mathematical Biosciences and Engineering},
	number = {5},
	pages = {4905},
	publisher = {NIH Public Access},
	title = {Identifying the spatial and temporal dynamics of molecularly-distinct glioblastoma sub-populations},
	volume = {17},
	year = {2020}}

@article{painter2002volume,
	author = {Painter, K. J. and Hillen, T.},
	journal = {Canadian Applied Mathematics Quarterly},
	number = {4},
	pages = {501--543},
	title = {Volume-filling and quorum-sensing in models for chemosensitive movement},
	volume = {10},
	year = {2002}}

@book{wiggins2003introduction,
	author = {Wiggins, S. and Golubitsky, M.},
	number = {3},
	publisher = {Springer},
	title = {Introduction to applied nonlinear dynamical systems and chaos},
	volume = {2},
	year = {2003}}

@article{perumpanani1999extracellular,
	author = {Perumpanani, A. J. and Byrne, H. M.},
	journal = {European Journal of Cancer},
	number = {8},
	pages = {1274--1280},
	publisher = {Elsevier},
	title = {Extracellular matrix concentration exerts selection pressure on invasive cells},
	volume = {35},
	year = {1999}}

@article{gerlee2016travelling,
	author = {Gerlee, P. and Nelander, S.},
	journal = {Mathematical Biosciences},
	pages = {75--81},
	publisher = {Elsevier},
	title = {Travelling wave analysis of a mathematical model of glioblastoma growth},
	volume = {276},
	year = {2016}}

@article{okubo1989spatial,
	author = {Okubo, A. and Maini, P. K. and Williamson, M. H. and Murray, J. D.},
	journal = {Proceedings of the Royal Society of London. B. Biological Sciences},
	number = {1291},
	pages = {113--125},
	publisher = {The Royal Society London},
	title = {On the spatial spread of the grey squirrel in Britain},
	volume = {238},
	year = {1989}}

@article{maini2004traveling,
	author = {Maini, P. K. and McElwain, D. L. S. and Leavesley, D. I.},
	journal = {Tissue Engineering},
	number = {3-4},
	pages = {475--482},
	publisher = {Mary Ann Liebert, Inc.},
	title = {Traveling wave model to interpret a wound-healing cell migration assay for human peritoneal mesothelial cells},
	volume = {10},
	year = {2004}}

@book{kot2001elements,
	author = {Kot, M.},
	publisher = {Cambridge University Press},
	title = {Elements of mathematical ecology},
	year = {2001}}

@article{painter2003modelling,
	author = {Painter, K. J. and Sherratt, J. A.},
	journal = {Journal of Theoretical Biology},
	number = {3},
	pages = {327--339},
	publisher = {Elsevier},
	title = {Modelling the movement of interacting cell populations},
	volume = {225},
	year = {2003}}

@article{anderson2009microenvironment,
  title={Microenvironment driven invasion: A multiscale multimodel investigation},
  author={Anderson, A. R. A. and Rejniak, K. A. and Gerlee, P. and Quaranta, V.},
  journal={Journal of Mathematical Biology},
  volume={58},
  pages={579--624},
  year={2009},
  publisher={Springer}
}

@article{malik2020impact,
  title={The impact of elastic deformations of the extracellular matrix on cell migration},
  author={Malik, A. A. and Wennberg, B. and Gerlee, P.},
  journal={Bulletin of Mathematical Biology},
  volume={82},
  pages={1--19},
  year={2020},
  publisher={Springer}
}

@article{tam2018nutrient,
  title={Nutrient-limited growth with non-linear cell diffusion as a mechanism for floral pattern formation in yeast biofilms},
  author={Tam, A. and Green, J. E. F. and Balasuriya, S. and Tek, E. L. and Gardner, J. M. and Sundstrom, J. F. and Jiranek, V. and Binder, B. J.},
  journal={Journal of theoretical biology},
  volume={448},
  pages={122--141},
  year={2018},
  publisher={Elsevier}
}

@article{kolmogorov1937study,
	author = {Kolmogorov, A. N. and Petrovskii, I. and Piskunov, N. S.},
	journal = {Moscow University Biological Sciences Bulletin},
	number = {6},
	pages = {1--25},
	title = {A study of the equation of diffusion with increase in the quantity of matter, and its application to a biological problem},
	volume = {1},
	year = {1937}}

@article{taylor2016coupling,
	author = {Taylor, P. R. and Baker, R. E. and Simpson, M. J. and Yates, C. A.},
	journal = {Journal of the Royal Society Interface},
	number = {120},
	pages = {20160336},
	publisher = {The Royal Society},
	title = {Coupling volume-excluding compartment-based models of diffusion at different scales: Voronoi and pseudo-compartment approaches},
	volume = {13},
	year = {2016}}

@article{browning2019bayesian,
	author = {Browning, A. P. and Haridas, P. and Simpson, M. J.},
	journal = {Bulletin of Mathematical Biology},
	number = {3},
	pages = {676--698},
	publisher = {Springer},
	title = {A Bayesian sequential learning framework to parameterise continuum models of melanoma invasion into human skin},
	volume = {81},
	year = {2019}}

@article{simpson2009diffusing,
	author = {Simpson, M. J. and Hughes, B. D. and Landman, K. A.},
	journal = {Australasian Journal of Engineering Education},
	number = {2},
	pages = {59--68},
	publisher = {Taylor \& Francis},
	title = {Diffusing populations: Ghosts or folks?},
	volume = {15},
	year = {2009}}

@article{bruna2012diffusion,
	author = {Bruna, M. and Chapman, J. S.},
	journal = {The Journal of Chemical Physics},
	number = {20},
	pages = {204116},
	publisher = {American Institute of Physics},
	title = {Diffusion of multiple species with excluded-volume effects},
	volume = {137},
	year = {2012}}

@article{simpson2009multi,
	author = {Simpson, M. J. and Landman, K. A. and Hughes, B. D.},
	journal = {Physica A: Statistical Mechanics and its Applications},
	number = {4},
	pages = {399--406},
	publisher = {Elsevier},
	title = {Multi-species simple exclusion processes},
	volume = {388},
	year = {2009}}

@article{canosa1973nonlinear,
	author = {Canosa, J.},
	journal = {IBM Journal of Research and Development},
	number = {4},
	pages = {307--313},
	publisher = {IBM},
	title = {On a nonlinear diffusion equation describing population growth},
	volume = {17},
	year = {1973}}

@article{el2021travelling,
	author = {El-Hachem, M. and McCue, S. W. and Simpson, M. J.},
	journal = {Physica D: Nonlinear Phenomena},
	pages = {133026},
	publisher = {Elsevier},
	title = {Travelling wave analysis of cellular invasion into surrounding tissues},
	volume = {428},
	year = {2021}}

@article{dallon_mathematical_1999,
	author = {Dallon, J. C. and Sherratt, J. A. and Maini, P. K.},
	%doi = {10.1006/jtbi.1999.0971},
	%issn = {00225193},
	journal = {Journal of Theoretical Biology},
	%language = {en},
	%month = aug,
	number = {4},
	pages = {449--471},
	shorttitle = {Mathematical modelling of extracellular matrix dynamics using discrete cells},
	title = {Mathematical modelling of extracellular matrix dynamics using discrete cells: Fiber orientation and tissue regeneration},
	%url = {https://linkinghub.elsevier.com/retrieve/pii/S0022519399909712},
	%urldate = {2022-04-06},
	volume = {199},
	year = {1999},
	%Bdsk-Url-1 = {https://linkinghub.elsevier.com/retrieve/pii/S0022519399909712},
	%Bdsk-Url-2 = {https://doi.org/10.1006/jtbi.1999.0971}
}

@article{perumpanani_extracellular_1998,
	abstract = {Cells use a combination of changes in adhesion, proteolysis and motility (directed and random) during the process of migration. Proteolysis of the extracellular matrix (ECM) results in thecreation of haptotactic gradients which cells use to move in a directed fashion. The proteolytic creation of these gradients also results in the production of digested fragments of ECM. In this study we show that in the human fibrosarcoma cell line HT1080, matrix metalloproteinase-2(MMP-2)-digested fragments of fibronectin exert a chemotactic pull stronger than that of undigested fibronectin. During invasion, this gradient of ECM fragments is established in the wake of an invading cell, running counter to the direction of invasion. The resultant chemotactic pull is anti-invasive, contrary to the traditional view of the role of chemotaxis in invasion. Uncontrolled ECM degradation by high concentrations of MMP can thus result in steep gradients of ECM fragments, which run against the direction of invasion. Consequently, the invasive potential of a cell depends on MMP production in a biphasic mannerimplying that MMP inhibitors will upregulate invasion in high-MMPexpressing cells. Hence the therapeutic use of protease inhibitors against tumours expressing high levels of MMP could produce an augmentation of invasion.},
	author = {Perumpanani, A. J. and Simmons, D. L. and Gearing, A. J. H. and Miller, K. M. and Ward, G. and Norbury, J. and Schneemann, M. and Sherratt, J. A.},
	%doi = {10.1098/rspb.1998.0582},
	%issn = {0962-8452},
	journal = {Proceedings of the Royal Society B: Biological Sciences},
	%month = dec,
	number = {1413},
	pages = {2347},
	%pmcid = {PMC1689540},
	%pmid = {null},
	title = {Extracellular matrix-mediated chemotaxis can impede cell migration},
	%url = {https://www.ncbi.nlm.nih.gov/pmc/articles/PMC1689540/},
	%urldate = {2022-03-04},
	volume = {265},
	year = {1998},
	%Bdsk-Url-1 = {https://www.ncbi.nlm.nih.gov/pmc/articles/PMC1689540/},
	%Bdsk-Url-2 = {https://doi.org/10.1098/rspb.1998.0582}
}

@article{strobl_mix_2020,
	abstract = {Invasion of healthy tissue is a defining feature of malignant tumours. Traditionally, invasion is thought to be driven by cells that have acquired all the necessary traits to overcome the range of biological and physical defences employed by the body. However, in light of the ever-increasing evidence for geno- and phenotypic intratumour heterogeneity, an alternative hypothesis presents itself: could invasion be driven by a collection of cells with distinct traits that together facilitate the invasion process? In this paper, we use a mathematical model to assess the feasibility of this hypothesis in the context of acid-mediated invasion. We assume tumour expansion is obstructed by stroma which inhibits growth and extra-cellular matrix (ECM) which blocks cancer cell movement. Further, we assume that there are two types of cancer cells: (i) a glycolytic phenotype which produces acid that kills stromal cells and (ii) a matrix-degrading phenotype that locally remodels the ECM. We extend the Gatenby--Gawlinski reaction--diffusion model to derive a system of five coupled reaction--diffusion equations to describe the resulting invasion process. We characterise the spatially homogeneous steady states and carry out a simulation study in one spatial dimension to determine how the tumour develops as we vary the strength of competition between the two phenotypes. We find that overall tumour growth is most extensive when both cell types can stably coexist, since this allows the cells to locally mix and benefit most from the combination of traits. In contrast, when inter-species competition exceeds intra-species competition the populations spatially separate and invasion arrests either: (i) rapidly (matrix-degraders dominate) or (ii) slowly (acid-producers dominate). Overall, our work demonstrates that the spatial and ecological relationship between a heterogeneous population of tumour cells is a key factor in determining their ability to cooperate. Specifically, we predict that tumours in which different phenotypes coexist stably are more invasive than tumours in which phenotypes are spatially separated.},
	author = {Strobl, M. A. R. and Krause, A. L. and Damaghi, M. and Gillies, R. and Anderson, A. R. A. and Maini, P. K.},
	%doi = {10.1007/s11538-019-00675-0},
	%issn = {0092-8240, 1522-9602},
	journal = {Bulletin of Mathematical Biology},
	%language = {en},
	%month = jan,
	number = {1},
	pages = {15},
	shorttitle = {Mix and {Match}},
	title = {Mix and match: {Phenotypic} coexistence as a key facilitator of cancer invasion},
	%url = {http://link.springer.com/10.1007/s11538-019-00675-0},
	%urldate = {2022-02-28},
	volume = {82},
	year = {2020},
	%Bdsk-Url-1 = {http://link.springer.com/10.1007/s11538-019-00675-0},
	%Bdsk-Url-2 = {https://doi.org/10.1007/s11538-019-00675-0}
}

@article{fisher_wave_1937,
	author = {Fisher, R. A.},
	%doi = {10.1111/j.1469-1809.1937.tb02153.x},
	%issn = {20501420},
	journal = {Annals of Eugenics},
	%language = {en},
	%month = jun,
	number = {4},
	pages = {355--369},
	title = {{The} wave of advance of advantageous genes},
	%url = {https://onlinelibrary.wiley.com/doi/10.1111/j.1469-1809.1937.tb02153.x},
	%urldate = {2021-12-20},
	volume = {7},
	year = {1937},
	%Bdsk-Url-1 = {https://onlinelibrary.wiley.com/doi/10.1111/j.1469-1809.1937.tb02153.x},
	%Bdsk-Url-2 = {https://doi.org/10.1111/j.1469-1809.1937.tb02153.x}
}

@book{morton_numerical_2005,
	abstract = {This is the 2005 second edition of a highly successful and well-respected textbook on the numerical techniques used to solve partial differential equations arising from mathematical models in science, engineering and other fields. The authors maintain an emphasis on finite difference methods for simple  but representative examples of parabolic, hyperbolic and elliptic equations from the first edition. However this is augmented by new sections on finite volume methods, modified equation analysis, symplectic integration schemes, convection-diffusion problems, multigrid, and conjugate gradient methods; and several sections, including that on the energy method of analysis, have been extensively rewritten to reflect modern developments. Already an excellent choice for students and teachers in mathematics, engineering and computer science departments, the revised text includes more latest theoretical and industrial developments.},
	%address = {Cambridge},
	author = {Morton, K. W. and Mayers, D. F.},
	%doi = {10.1017/CBO9780511812248},
	edition = {2},
	%isbn = {978-0-521-60793-3},
	publisher = {Cambridge University Press},
	shorttitle = {Numerical solution of partial differential equations},
	title = {Numerical solution of partial differential equations: {An} introduction},
	%url = {https://www.cambridge.org/core/books/numerical-solution-of-partial-differential-equations/EB8E5037C4A49F78D91C0AF7EE4CC7FA},
	%urldate = {2021-12-15},
	year = {2005},
	%Bdsk-Url-1 = {https://www.cambridge.org/core/books/numerical-solution-of-partial-differential-equations/EB8E5037C4A49F78D91C0AF7EE4CC7FA},
	%Bdsk-Url-2 = {https://doi.org/10.1017/CBO9780511812248}
}

@book{murray2001mathematical,
  title={Mathematical biology II: {S}patial models and biomedical applications},
  author={Murray, J. D.},
  volume={3},
  year={2001},
  publisher={Springer New York}
}

@book{lam2022introduction, 
 title = {Introduction to reaction-diffusion equations: {T}heory and applications to spatial ecology and evolutionary biology}, 
 author={Lam, K.-Y. and Lou, Y.},
 publisher={Springer},
 year={2022}
}

@book{murray2002mathematical,
  title={Mathematical biology I: An introduction},
  author={Murray, J. D.},
  year={2002},
  publisher={Springer}
}

@article{colson_travelling-wave_2021,
	author = {Colson, C. and S{\'a}nchez-Gardu{\~n}o, F. and Byrne, H. M. and Maini, P. K. and Lorenzi, T.},
	journal = {Proceedings of the Royal Society A},
	number = {2256},
	pages = {20210593},
	publisher = {The Royal Society},
	title = {Travelling-wave analysis of a model of tumour invasion with degenerate, cross-dependent diffusion},
	volume = {477},
	year = {2021}}


\begin{appendices}
\appendixpage
\section{Main results of numerical simulations for $m_0=1$}\label{m0_1}
As demonstrated in Figure~\ref{fig:cvsdmvsm0}, when $m_0=1$, the system~\eqref{NDeqn2_u}-\eqref{NDeqn2_m} subject to the initial conditions~\eqref{hIC_u}-\eqref{hIC_m} does not permit travelling wave solutions.
To investigate this further, we simulate the system~\eqref{NDeqn2_u}-\eqref{NDeqn2_m} subject to different initial conditions. 
{In every case, we consider the initial condition for the ECM density, $m$, given by
 \begin{align}
    m(x,0)= \begin{cases} 
    m_0-u(x,0), \qquad &$if$ \qquad m_0> 1-\gamma, \\
    m_0, \qquad &$if$ \qquad m_0\leq1-\gamma, \label{oldIC_m}
    \end{cases} 
    \end{align}
with $0\le\gamma\leq1$, which depends on the initial cell density, $u(x,0)$. 
To explore the behaviours observed at $m_0=1$, we consider two different options for $u(x,0)$ in this appendix. 
First, we consider the compactly supported initial condition
\begin{equation}
    u(x,0)=\begin{cases}\gamma\big(1-\tanh(\frac{x}{\epsilon})\big), \qquad &$if$ \qquad \gamma\big(1-\tanh(\frac{x}{\epsilon})\big)\geq \xi , \\ 0, \qquad &$if$ \qquad \gamma\big(1-\tanh(\frac{x}{\epsilon})\big)<\xi, \end{cases}\label{oldIC_u} 
\end{equation}
and alternatively, the following non-compactly supported initial conditions, as used in~\cite{el2021travelling},
\begin{align}
    u(x,0) &= \begin{cases} 
    \gamma, \qquad &x<\beta, \\
    \gamma \exp\{-a (x-\beta)\}, \qquad &x\geq\beta. \label{oldIC2_u}
    \end{cases} 
\end{align}}
Here, $\gamma\in[0,1]$ represents the maximum cell density at $t=0$ and $m_0\in[0,1]$ corresponds to the uninvaded density of ECM. Moreover, in the definition given by Equation~\eqref{oldIC_u}, the parameter $\xi \in(0,1]$ is used to control the tolerance below which the cell density can be assumed, on a first approximation, to be zero, and $\epsilon>0$ represents the initial width of the cell density profile.
Finally, in the definition given by Equation~\eqref{oldIC2_u}, the parameter $\beta\in\mathbb{R}$ is used to define a region where the cell density is initially constant and equal to $\gamma\in[0,1]$, while the parameter $a>0$ is used to prescribe the lengthscale over which the cell density profile decays. 
We note that, since $\gamma,\, m_0\in[0,1]$, the initial conditions~\eqref{oldIC_m},~\eqref{oldIC_u}~and~\eqref{oldIC2_u} are such that the total density of cells and ECM at $t=0$ does not locally exceed the extreme value $1$, which corresponds to complete local saturation, i.e. $u(x,0) + m(x,0) \leq 1$ for all $x \in [0,L]$. We also note that when $m_0=0$ the initial condition~\eqref{oldIC_m} reduces to the trivial initial condition $m(x,0) \equiv 0$.

\begin{figure}
    \centering
    (a)\includegraphics[scale=0.6]{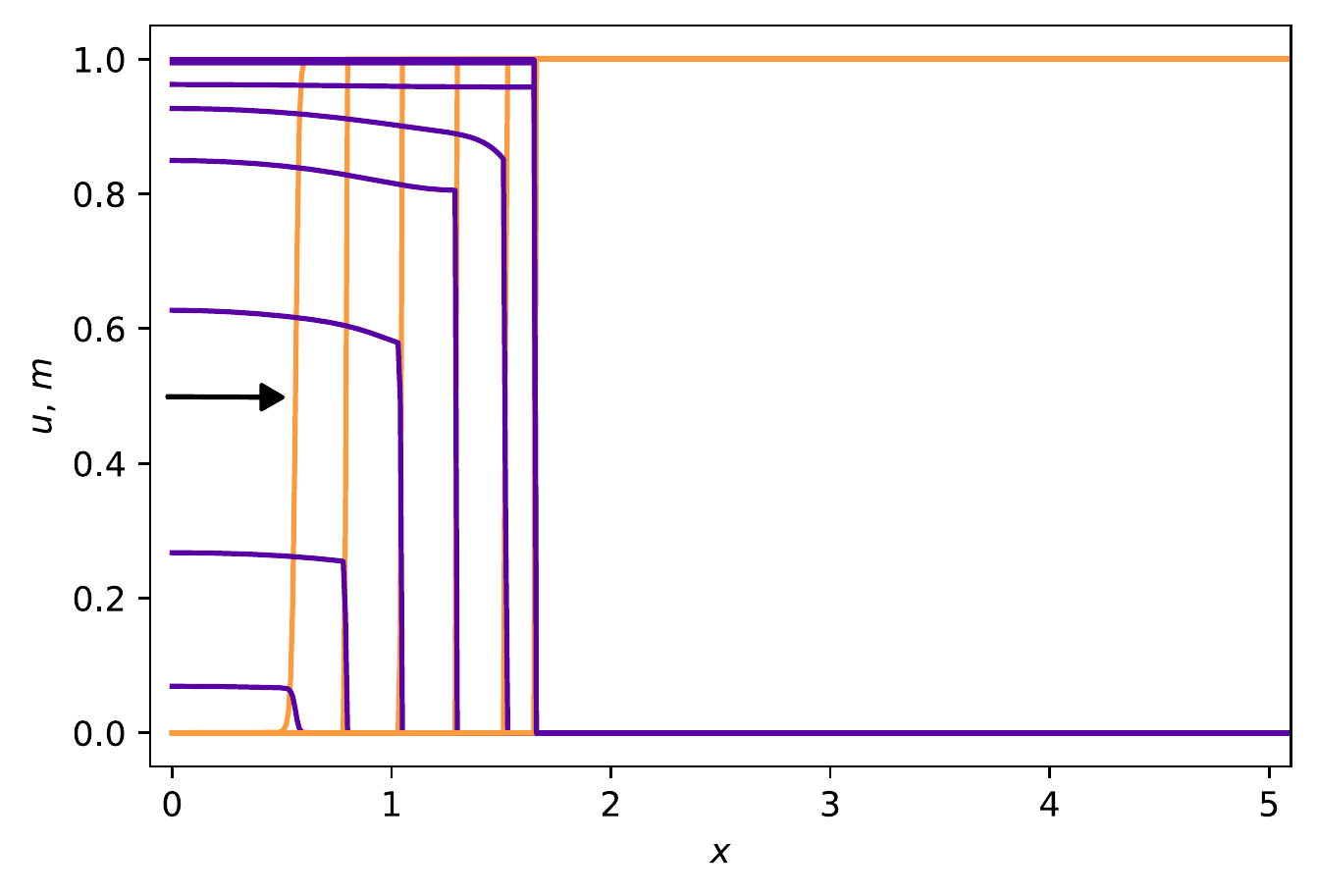}
    (b)\includegraphics[scale=0.6]{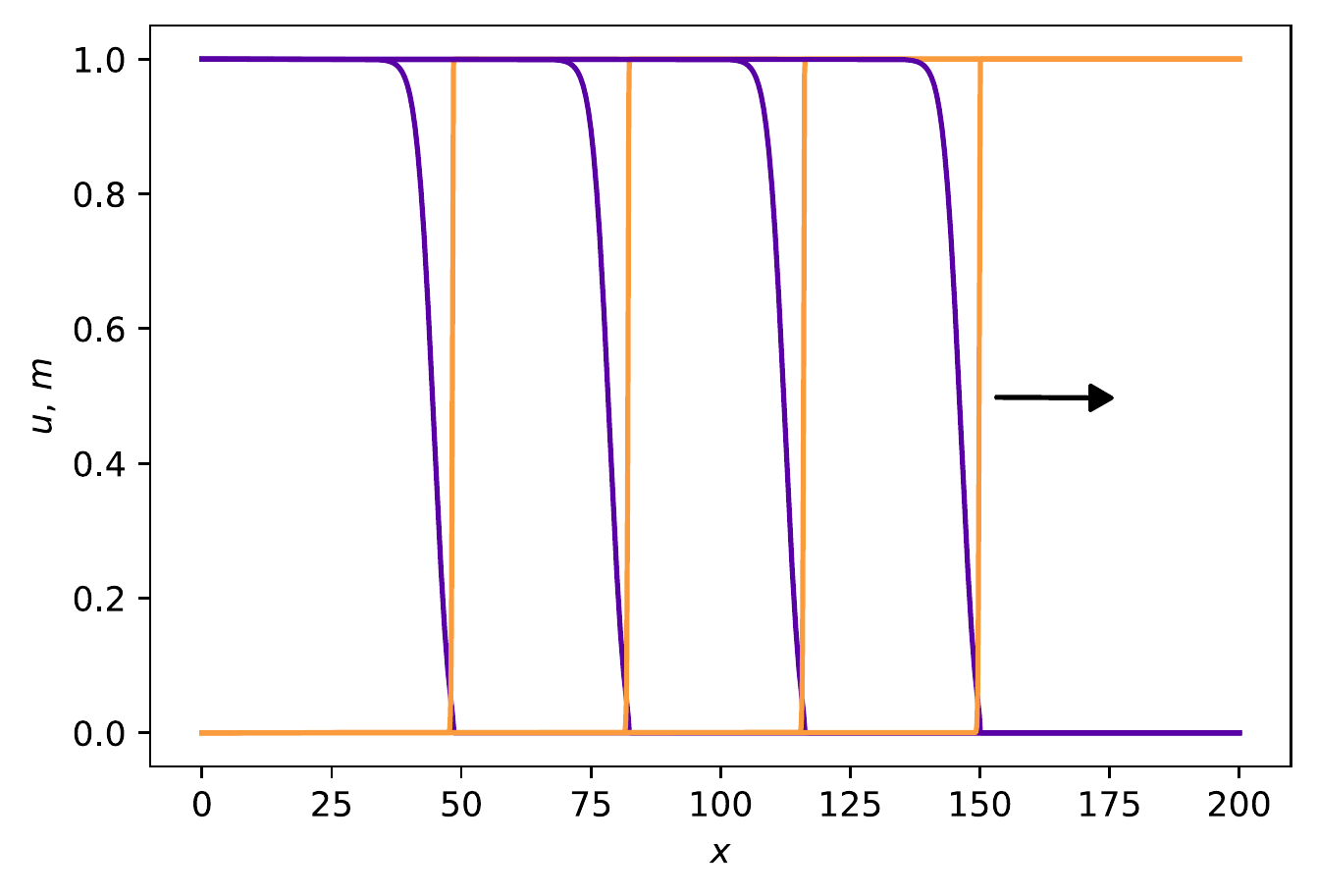}
    \caption{Numerical solutions to the system~\eqref{NDeqn2_u}-\eqref{NDeqn2_m} subject to the initial conditions~\eqref{oldIC_m}~and~\eqref{oldIC_u} (panel (a)) or~\eqref{oldIC_m}~and~\eqref{oldIC2_u} (panel (b)), for $m_0=1$ and $\lambda=250$.  Cell densities are shown in purple and ECM densities in orange at times $t=2,4,6,8,10,12, 14, 16$ (from left to right) in panel (a) and times $t=25,50,75,100$ (from left to right) in panel (b). Note that the axis in the plot in panel (a) are zoomed in on $x\in[0,5]$ to display the initial behaviour in the transient region before invasion stops. Further specifics of the parameter values and the numerical methods used for simulation can be found in Appendix~\ref{appNS}.}
    \label{fig:CSvsNCS}
\end{figure}

The numerical results in Figure~\ref{fig:CSvsNCS}(a), which complement the results summarised by Figure~\ref{fig:cvsdmvsm0}, show that when $m_0 =~1$ the system~\eqref{NDeqn2_u}-\eqref{NDeqn2_m} subject to the initial conditions with compactly support cell density~\eqref{oldIC_u} cannot sustain travelling wave solutions. 
On the other hand, the numerical results in Figure~\ref{fig:CSvsNCS}(b) demonstrate that travelling wave solutions can be sustained in the case where non-compactly supported initial conditions~\eqref{oldIC2_u} in $u$ are considered. 

This result is a consequence of the volume-filling effects of cells. 
By considering an initial condition where $m(x,0)=~1$ ahead of the invading population, due to volume-filling, the invading population is unable to penetrate the region where $u=0,\,m=1$. 
This agrees with the agent-based description, since cells are only able to degrade ECM in the same lattice site.

As such, for the model~\eqref{NDeqn2_u}-\eqref{NDeqn2_m}, whenever $m_0=1$ and there are compactly supported initial conditions in $u$, invasion is entirely prevented beyond a point $x^{*}$, that is the smallest $x$ such that $u({x},0)=0$ for all $x\geq x^{*}$.
This result is starkly different to simpler models in the literature that do not include volume-filling effects of cells and ECM, such as  \cite{el2021travelling, colson_travelling-wave_2021}, where the total density of cells and ECM is not bounded above and cells can invade into a region where~$u=0,\,m=1$, and thus exhibit travelling wave solutions.

\section{Numerical methods} \label{appNS}

Equations~\eqref{NDeqn2_u}-\eqref{NDeqn2_m} are solved numerically subject to no flux boundary conditions~\eqref{NFBC} in $u$ at $x=0$ and $x=L$ using the method of lines on the 1D spatial domain $[0,L]$ where $L>0$ is chosen to be sufficiently large to remove boundary effects. In most cases, we take $L=200$. 
The spatial domain is uniformly discretised with spacing $\Delta=0.1$ between each of the $i=1,\,\dots , \,I$ spatial points, and the following discretisation is used \cite{strobl_mix_2020}: 
\begin{equation}
    \frac{\partial}{\partial x}\bigg[D\frac{\partial a}{\partial x}\bigg]_{i}\approx \frac{1}{2\Delta^2}\bigg[(D_{i-1}+D_{i})a_{i-1}-(D_{i-1}+2D_{i}+D_{i+1})a_i+(D_i+D_{i+1})a_{i+1}\bigg],
\end{equation}
where $a_{i}$ represents the value of $a$ at the spatial point $i.$ 
For the model~\eqref{NDeqn2_u}-\eqref{NDeqn2_m}, we use this discretisation twice, with $D=(1-m), \, a=u$ and for the second term in the flux as $D=u, \, a=m$. 
Equations~\eqref{NDeqn2_u}-\eqref{NDeqn2_m} can then be rewritten as a system of $2I$ ordinary differential equations given by: 
\begin{align}
    \frac{\mathrm{d} u_i}{\mathrm{d} t} &= \frac{1}{2\Delta^2}\bigg[u_{i-1}(1-m_i)+u_i(m_{i+1}+m_{i-1}-2)+u_{i+1}(1-m_i)\bigg]+u_i(1-u_i-m_i), \label{u1disc}\\
    \frac{\mathrm{d} m_i}{\mathrm{d} t} &= -\lambda m_i u_i, \label{mdisc}
\end{align}
for $1\leq i \leq I-1.$
 To implement the boundary conditions, we introduce the ghost points $x_{-1}$ and $x_{I+1}$ \cite{morton_numerical_2005} and set 
\begin{equation}
    u_0(t)=u_{-1}(t), \qquad u_{I+1}(t)=u_{I}(t), \qquad \forall t\geq0,
\end{equation}
so that 
\begin{align}
    \frac{\mathrm{d} u_0}{\mathrm{d} t}&=2(u_1-u_0)+u_0(1-u_0-m_0), \label{nofluxdisc} \\
    \frac{\mathrm{d} u_I}{\mathrm{d} t}&=2(u_{I-1}-u_I)+u_I(1-u_I-m_I). \label{nofluxdisc2}
\end{align}
We solve the system of equations~\eqref{u1disc}-\eqref{mdisc} and~\eqref{nofluxdisc}-\eqref{nofluxdisc2} using the built-in Python solver \text{scipy.integrate.solve\_ivp} with the explicit Runge-Kutta integration method of order 5 and time step $\tau=1$.  Convergence checks were completed by considering a range of tolerances, time and spatial steps,  to ensure that the parameters used for simulations produced solutions within the second order error associated with the numerical scheme.

For the simulations of the PDE systems in this work,  we consider compactly supported initial conditions~\eqref{hIC_u}-\eqref{hIC_m} with $\alpha=1$. 
In Appendix~\ref{m0_1} we use $\xi=10^{-7}, \, \gamma=0.1$ and $\epsilon=1$ when considering compactly supported initial conditions~\eqref{oldIC_m}~and~\eqref{oldIC_u}, and $a=0.1, \, \gamma=0.1$ and $\beta=10$ for non-compactly supported initial conditions~\eqref{oldIC_m}~and~\eqref{oldIC2_u}. Varying these parameters reproduces the behaviours observed in \cite{el2021travelling}.

In Figure~\ref{fig:cODEpp}, we show the results of numerically solving Equations~\eqref{stdTW3_u}-\eqref{stdTW3_v} with the initial condition $(U, V, M)=(0.9, -0.01, 0.01)$ for $c=1,\, 2(1-m_0), \,3$ with time step $\tau=0.01$ and final time $t=100$ using Python's built-in stiff solver \text{scipy.integrate.ODE} with tolerance $10^{-15}$ and order $5$.

\section{Comparison to other models in the literature}\label{APPcomp}
This study focuses on the impact of introducing volume-filling effects of cells and ECM to a model of cell invasion into ECM. 
There are a number of PDE model simplifications in the literature, including the following model, proposed as a minimal model for tumour growth into ECM in \cite{colson_travelling-wave_2021}:
\begin{align}
    \frac{\partial u}{\partial t}&=  \frac{\partial}{\partial x}\bigg[ (1- m)\frac{\partial u}{\partial x}\bigg]+ u(1-u),  \label{Colson_u} \\
    \frac{\partial m}{\partial t}&=-\lambda m u, \label{Colson_m}
\end{align}
that assumes cell motility to be impacted by the presence of surrounding ECM only and cell proliferation impacted only by other cells, that is, the resource limiting cell proliferation is not space. 
Another similar model is presented in \cite{browning2019bayesian} to describe melanoma growth into skin and it is subsequently analysed in \cite{el2021travelling}.
The model can be interpreted to assume that cell motility is decreased by ECM, and that cell proliferation is impacted by both other cells and ECM: 
\begin{align}
    \frac{\partial u}{\partial t}&=  \frac{\partial}{\partial x}\bigg[ (1- m)\frac{\partial u}{\partial x}\bigg]+ u(1-u-m),  \label{Browning_u} \\
    \frac{\partial m}{\partial t}&=-\lambda m u. \label{Browning_m}
\end{align}
The model variables and parameters are interpreted in the same way as in the model presented in this work~\eqref{NDeqn2_u}-\eqref{NDeqn2_m}. 

\begin{figure}[h!]
\hspace*{-.3cm} 
\includegraphics[scale=0.375]{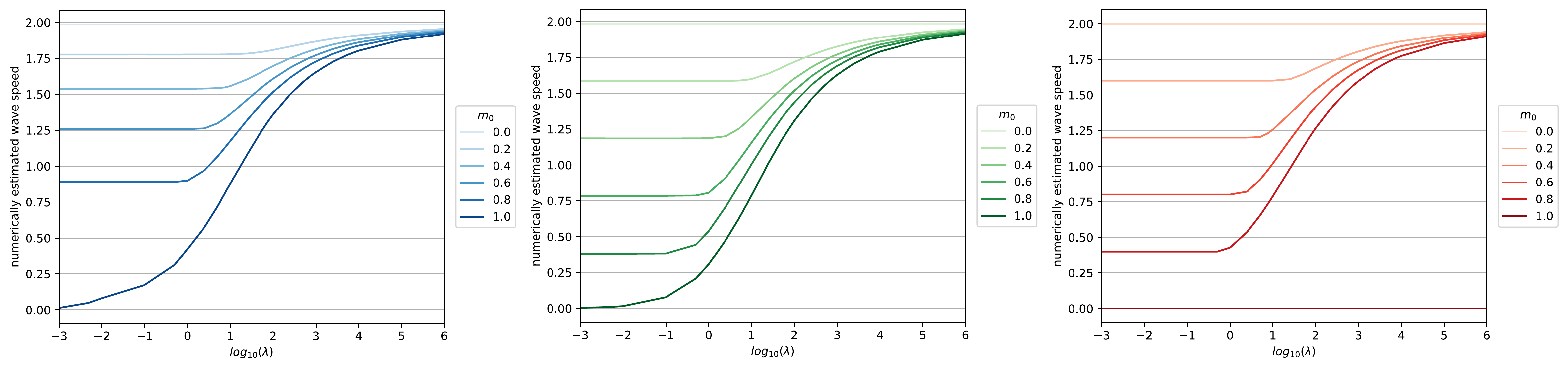}
\caption{The relationship between the numerically estimated speed of travelling wave solutions to the system~\eqref{Colson_u}-\eqref{Colson_m} on the left (blue), \eqref{Browning_u}-\eqref{Browning_m} in the middle (green) and \eqref{NDeqn2_u}-\eqref{NDeqn2_m} on the right (red), subject to the initial conditions \eqref{hIC_u}-\eqref{hIC_m}. The numerically estimated travelling wave speed is obtained by tracing the point $X(t)$ such that $u(X(t), t)=0.1.$ Further specifics of the parameter values and the numerical methods used can be found in Appendix~\ref{appNS}.}
\label{fig:comparisonc}
\end{figure}

We are particularly interested in comparing the population-level behaviours of the PDE model for cell invasion into ECM presented in this work, which incorporates volume-filling effects into both diffusion and proliferation of cells,  to the simpler models without these volume-filling effects, presented in the literature. 
By looking at Figure~\ref{fig:comparisonc}, we can draw the following conclusions: all three models produce travelling wave solutions with a speed $c\geq c_{\text{min}}$, where $c_{\text{min}}$ is the minimum speed predicted by standard travelling wave analysis. 
In fact, all of these speeds are dependent on both the initial density of ECM ahead of the wave, $m_0$, and the rescaled ECM degradation rate $\lambda$. 
The two models with the same reaction (growth) terms, depending on both cell and ECM preventing growth, predict the same travelling wave speed $c_{\text{min}}=2(1-m_0)$, that is achieved numerically for $\lambda\to0^{+}$. 
However, the model~\eqref{Colson_u}-\eqref{Colson_m} presented in \cite{colson_travelling-wave_2021} predicts a speed  $c_{\text{min}}=2\sqrt{1-m_0},$ that is also revealed for $\lambda\to0^{+}.$
As a result, the behaviours observed in these models for small rescaled ECM degradation rates $\lambda$ can be reproduced by studying a FKPP model~\eqref{FKPP} with the appropriate parameters. 
In the same manner, by looking at Figure~\ref{fig:comparisonc}, it is clear that all three models produce travelling waves with speed $c\to2^{-}$ as $\lambda\to\infty$.
The behaviours observed here can be studied by considering the standard FKPP model~\eqref{FKPP1} with all parameters equal to unity. 
The FKPP model~\eqref{FKPP1} is also a suitable model simplification for all three systems when $m_0=0$.  

The transition between the two asymptotic regions is yet to be fully characterised for any of the models, but it is clear that $c$ is a monotonic, increasing function of $\lambda$ and $m_0$ for all of the models. 
The critical value above which $\lambda$ begins to influence the speed is similar across the models, but clearly depends on 
$m_0$ and takes larger values across the models as more volume-filling effects are taken into account. 
Following intuition, we also find that, in general, the speed of invasion is slower as volume-filling effects are considered to impact more aspects of cell behaviours (from left to right in Figure~\ref{fig:comparisonc}).

The most obvious difference between these results is that the model~\eqref{NDeqn2_u}-\eqref{NDeqn2_m} derived in this work does not permit travelling waves for compactly supported initial conditions in $u$ when $m_0=1$. This is a direct result of consistently including volume-filling effects across all the mechanisms of cell movement, such that there is always a maximum number of cells present at any point in space. The results match those of the model~\eqref{Browning_u}-\eqref{Browning_m} when non-compactly supported initial conditions are simulated, as presented in Appendix~\ref{m0_1}. 

As such, the model presented in \cite{browning2019bayesian} provides a good model simplification by which to study the qualitative properties of the solutions to~\eqref{NDeqn2_u}-\eqref{NDeqn2_m} across all parameter values when $m_0\in[0,1)$, with simplifications to the FKPP model also being appropriate as $\lambda\to0^{+}$ and $\lambda\to\infty$.

\section{Derivation of eigenvalues and eigenvectors}\label{APPeigs}
In this section, we derive the eigenvalues of the system of ordinary differential equations~\eqref{stdTW3_u}-\eqref{stdTW3_m}. 
This system has two equilibrium points  $\mathcal{S}_1=(1,0,0)$ and $\mathcal{S}_2=(0,0,m_0)$, at which we want to find eigenvalues.
We first find the Jacobian of the linearised system~\eqref{stdTW3_u}-\eqref{stdTW3_m}. 
To do this, we introduce the following combinations for simplicity
\begin{equation}
\nu = \frac{\lambda}{c}, \qquad N=\frac{1}{1-M}, \qquad W=MU,
\end{equation}
so that the Jacobian is given by
\begin{equation}
\mathbf{J}=
    \begin{pmatrix}
    0 & 1 & 0 \\
    N(M-1+2U-3\nu^2WU-\nu W) & -N\big(c+\nu W\big) & UN(1-\nu^2U^2-\nu V+N(U+M-1-\nu^2WU))-N^2(cV+\nu) \\
    \nu M & 0 & \nu U
    \end{pmatrix}.
\end{equation}
Then the Jacobian at $\mathcal{S}_2=(0,0,m_0)$ is 
\begin{equation}
\mathbf{J}_{(0,0,m_0)}=
    \begin{pmatrix}
    0 & 1 & 0 \\
    \dfrac{m_0-1}{1-m_0} & \dfrac{-c}{1-m_0} & 0 \\
    \dfrac{\lambda}{c}m_0 & 0 & 0
    \end{pmatrix},
\end{equation}
and the Jacobian at  $\mathcal{S}_1=(1,0,0)$ is
\begin{equation}
\mathbf{J}_{(1,0,0)}=
    \begin{pmatrix}
    0 & 1 & 0 \\
    1 & -c & 1-\bigg(\dfrac{\lambda}{c}\bigg)^2 \\
    0 & 0 & \dfrac{\lambda}{c}
    \end{pmatrix}.
\end{equation}
By looking for the solutions of $\text{det}|\mathbf{J}-\sigma \textbf{I}|=0$, where $\textbf{I}$ is the identity matrix, we can find the eigenvalues of these matrices, and calculate their corresponding eigenvectors.
As such, at $(1,0,0)$, the eigenvalues are: $\sigma_1={\lambda}/{c}$, $\sigma_{2,3}={(-c\pm\sqrt{c^2+4})}/{2},$ which have associated eigenvectors 
\begin{align}
\mathbf{v_1}&=\begin{pmatrix} \dfrac{c^2-\lambda^2}{c^2(\lambda-1)+\lambda^2}, & \dfrac{\lambda(c^2-\lambda^2)}{c(c^2(\lambda-1)+\lambda^2)}, & 1 \end{pmatrix}^{T}  , \\
\mathbf{v_{2,3}}&=\begin{pmatrix} \dfrac{c\pm\sqrt{c^2+4}}{2}, & 0, & 1 \end{pmatrix}^{T}  .
\end{align}
These indicate that $(1,0,0)$ is a three-dimensional, hyperbolic, unstable saddle point since it has one negative and two positive eigenvalues.

At $(0,0,m_0)$, $\text{det}|\mathbf{J}_{(0,0,m_0)}-\sigma \textbf{I}|=0$ gives eigenvalues $\sigma_1=0$, $\sigma_{2,3}=({-c\pm\sqrt{c^2-4(1-m_0)^2}})/{2(1-m_0)},$ showing that $(0,0,m_0)$ is a non-hyperbolic, stable steady state, since one of these eigenvalues has zero real part. 
If  $c^2<4(1-m_0)^2$, then we have a spiral at $(0,0,m_0),$ and otherwise, a stable node point.
The corresponding eigenvectors are
\begin{align}
\mathbf{w_1}&=\begin{pmatrix} 0, & 0, & 1 \end{pmatrix}^{T}  , \\
\mathbf{w_{2,3}}&=\begin{pmatrix} \dfrac{c(c\pm\sqrt{c^2-4(1-m_0)^2})}{2\lambda m_0(m_0-1)}, & \dfrac{c(c^2\pm c\sqrt{c^2-4(1-m_0)^2}-2(1-m_0)^2)}{2\lambda m_0 (1-m_0)^2}, & 1 \end{pmatrix} ^{T} .
\end{align} 

\section{Travelling wave profiles for $\lambda\to0^{+}$ and $\lambda\to\infty$}
\begin{figure}[h!]
    \centering
    \includegraphics[scale=0.575]{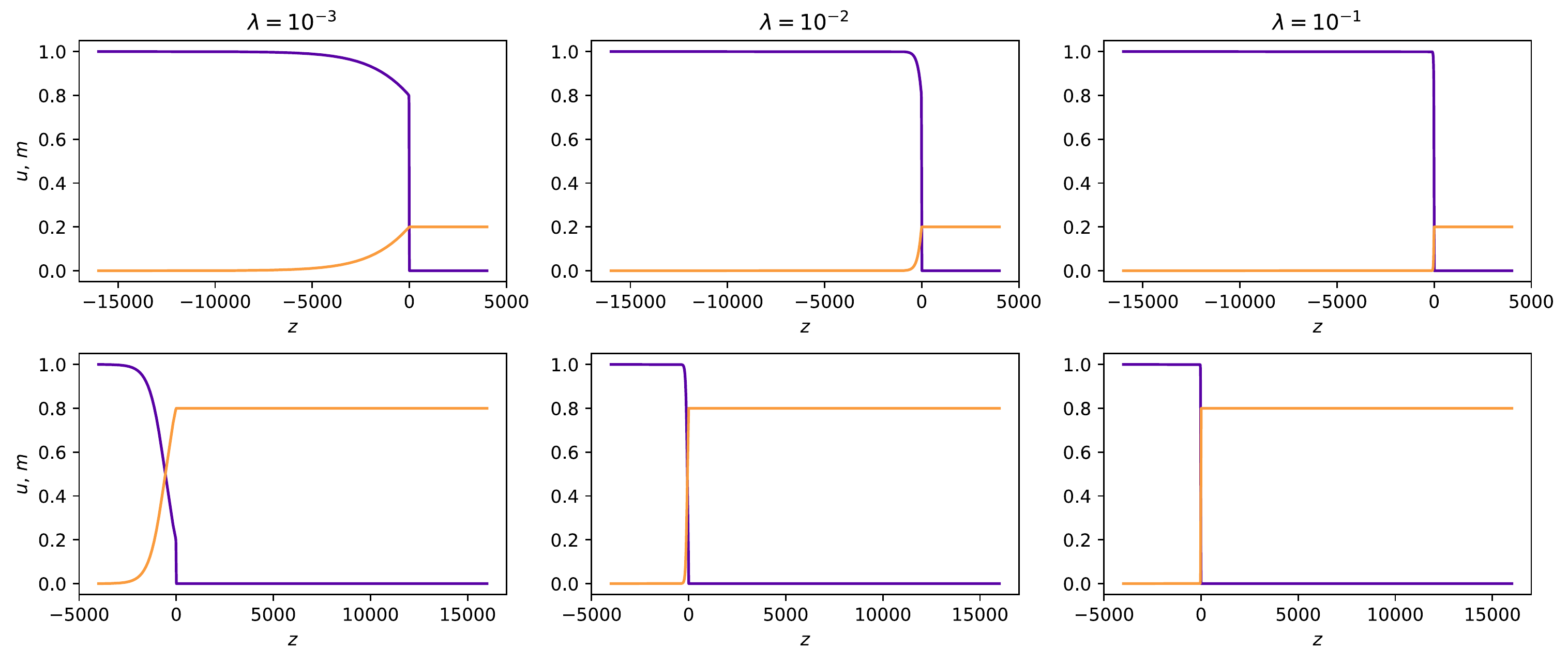}
    \caption{Travelling wave solutions of Equations~\eqref{NDeqn2_u}-\eqref{NDeqn2_m} subject to the initial conditions~\eqref{hIC_u}-\eqref{hIC_m}, for $m_0=0.2$ in the top row and $m_0=0.8$ in the bottom row, and for rescaled ECM degradation rates $\lambda=10^{-3}, \,10^{-2},\,10^{-1}.$ Cell densities are shown in purple and ECM densities in orange. Further specifics of the parameter values and the numerical methods used can be found in Appendix~\ref{appNS}.}
    \label{fig:small_lam_TW}
\end{figure}
\begin{figure}
    \centering
    \includegraphics[scale=0.575]{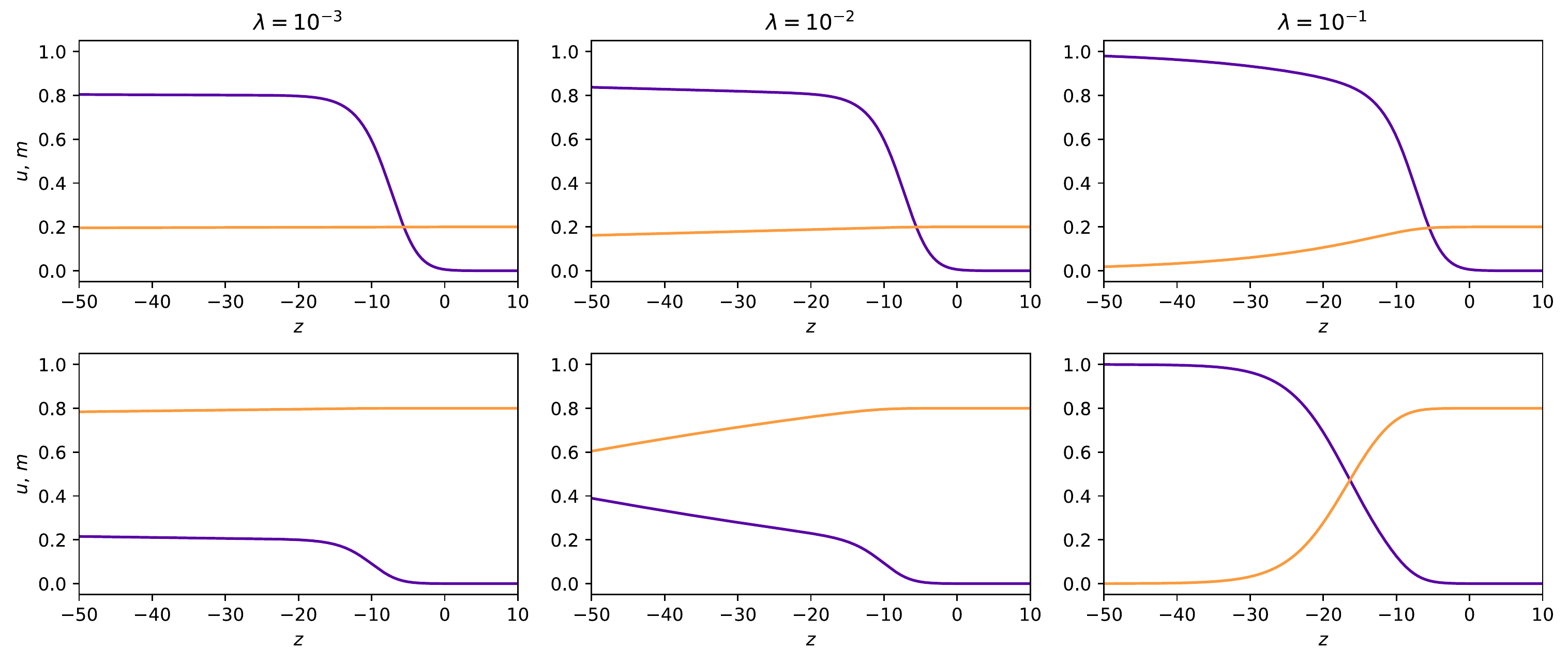}
    \caption{Travelling wave solutions of Equations~\eqref{NDeqn2_u}-\eqref{NDeqn2_m} subject to the initial conditions~\eqref{hIC_u}-\eqref{hIC_m}, for $m_0=0.2$ in the top row and $m_0=0.8$ in the bottom row, and for rescaled ECM degradation rates $\lambda=10^{-3}, \,10^{-2},\,10^{-1}.$ Cell densities are shown in purple and ECM densities in orange, zoomed in on the evolved travelling wave front, as shown in Figure~\ref{fig:small_lam_TW}. Further specifics of the parameter values and the numerical methods used can be found in Appendix~\ref{appNS}.}
    \label{fig:zoomed_small_lam}
\end{figure}
\begin{figure}[h!]
    \centering
    \includegraphics[scale=0.575]{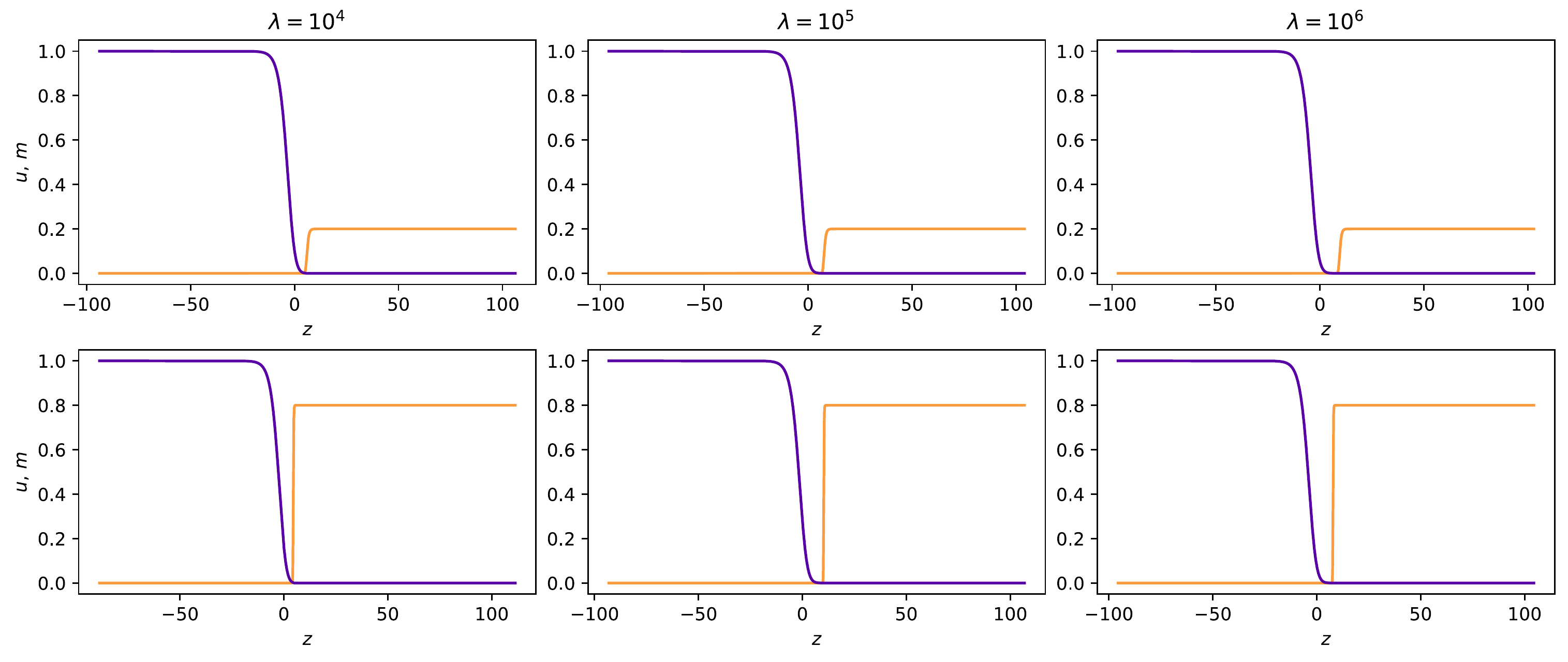}
    \caption{Travelling wave solutions of Equations~\eqref{NDeqn2_u}-\eqref{NDeqn2_m} subject to the initial conditions~\eqref{hIC_u}-\eqref{hIC_m}, for $m_0=0.2$ in the top row and $m_0=0.8$ in the bottom row, and for rescaled ECM degradation rates $\lambda=10^{4}, \,10^{5},\,10^{6}.$ Cell densities are shown in purple and ECM densities in orange. Further specifics of the parameter values and the numerical methods used can be found in Appendix~\ref{appNS}.}
    \label{fig:large_lam_TW}
\end{figure}

\end{appendices}

\end{document}